\documentclass[onecolumn,preprintnumbers,noshowpacs,noshowkeys,amsmath,amssymb,nofootinbib,superscriptaddress,floatfix]{revtex4}

\usepackage{subfigure}
\usepackage{bm}

\usepackage{amsmath,amssymb,bm}
\usepackage{epsfig}
\usepackage{graphicx}
\usepackage{pstricks,pst-node,pst-text,pst-3d}

\newcommand{\beeq}{\begin{eqnarray}}
\newcommand{\eeeq}{\end{eqnarray}}
\newcommand{\be}{\begin{equation}}
\newcommand{\ee}{\end{equation}}
\newcommand{\bea}{\begin{array}}
\newcommand{\eea}{\end{array}}
\newcommand{\beq}{\begin{equation}}
\newcommand{\eeq}{\end{equation}}

\newcommand{\eq}{\!&=&\!}

\newcommand{\qbar}{{\overline{q}}}

\newcommand{\intlim}{\int}

\newcommand{\sigmahat}{\sigma}
\newcommand{\rhat}{\hat{r}}
\newcommand{\rbo}{{\bf{r}}}
\newcommand{\Htilde}{{\tilde{H}}}
\newcommand{\sigmatilde}{{\tilde{\sigma}}}
\newcommand{\Omegatilde}{{\tilde{\Omega}}}

\newcommand{\alfaem}{\alpha_{em}}

\newcommand{\ebar}{{\left<e^2\right>}}
\newcommand{\mbf}[1]{\mbox{\boldmath $#1$}}
\newcommand{\bk}{\mbf{k}}
\newcommand{\bq}{\mbf{q}}

\begin{document}

\title{ \mbox{}\\ 
\bf Twist expansion of the nucleon structure functions, $\boldmath{F_2}$ and ${\boldmath F_L}$, \\ in the DGLAP improved saturation model}

\author{
Jochen Bartels$^{(a)}$,
Krzysztof~Golec-Biernat$^{(b,c)}$ and Leszek Motyka$^{(a,d)}$\\[3mm]
\textit{$^{a}$II Institute for Theoretical Physics, Hamburg University,\\
Luruper Chaussee 149, 22761 Hamburg, Germany}\\[1mm]
\textit{$^{b}$Institute of Physics, University of Rzesz\'ow, \\
Al.\ Rejtana 16 A, 35-959 Rzesz\'ow, Poland}\\[1mm]
\textit{$^{c}$Institute of Nuclear Physics Polish Academy of Sciences, \\
Radzikowskiego 152, 31-342 Krak\'{o}w, Poland}\\[1mm]
\textit{$^{d}$Institute of Physics, Jagiellonian University, \\
Reymonta 4, 30-059 Krak\'{o}w, Poland}\\[2mm]
}

\date{November 10th, 2009}

\begin{abstract}
\noindent
Higher twist effects in the deeply inelastic scattering are studied. We start with a short review of the theoretical results on higher twists in QCD.
Within the saturation model we perform a twist analysis of the nucleon structure functions $F_T$  and $F_L$ at small value of the Bjorken variable $x$. 
The parameters of the model are fitted to the HERA $F_2$ data, 
and we derive a prediction for the longitudinal structure function $F_L$. We conclude that for $F_L$ the higher twist corrections are sizable whereas for $F_2=F_T+F_L$ there is a nearly complete cancellation 
of twist-4 corrections in $F_T$ and $F_L$. We discuss a few consequences 
for future LHC measurements.        
\end{abstract}


\maketitle

\vspace{-11.5cm}
\hfill {\sc DESY 09-127}
\vspace{10.7cm}

\section{Introduction}
\label{sec:1}

\noindent
A deeper understanding of the transition region at low $Q^2$ and small $x$ 
in deep inelastic electron proton scattering has been one of the 
central tasks of HERA physics. Approaching the transition region from the 
perturbative side one expects to see the onset of corrections to the 
successful DGLAP description --- in particular those which belong to higher    twist operators in QCD. The twist expansion defines a systematic approach 
and, therefore, provides an attractive framework of investigating the 
region of validity of the leading twist DGLAP evolution equations.

The essentials of the theory of higher twist operators and their 
$Q^2$ evolution have been laid down several years ago, \cite{qBFKL,EFP1,EFP2}. First, a choice has to be made of a complete operator basis ~\cite{EFP1,EFP2},
and for their evolution~\cite{qBFKL} one needs to compute evolution kernels 
which, for partonic operators in leading order, reduce to $2 \to 2$ scattering 
kernels. The problem of mixing between different operators has also been 
addressed first in ~\cite{qBFKL}. In the small-$x$ region at HERA one expects 
the gluonic operators to be the most dominant ones; so far, a theoretical 
study of the evolution of twist-4 gluon operators is available only 
in the double logarithmic approximation ~\cite{TPV,DLA1,DLA2,DLA3}.
An extensive recent theoretical study of QCD evolution of the higher twist 
operators can be found in \cite{BMR}.

Numerical studies of the size of potential higher twist corrections 
~\cite{MRST,MR} indicate that twist-4 corrections to $F_2$ are small down 
to $Q^2 \sim 1$ GeV$^2$, $x\sim 10^{-4}$. A first theoretical analysis~\cite{Bartels:2000hv} applied to HERA data, however, has shown that the situation is more subtle, and from the smallness of twist-4 corrections to $F_2$ one cannot conclude that contributions of twist-4 operators are small. The simplest QCD diagrams contributing to the twist four gluon operator are shown in Fig.~\ref{fig1}: a quark loop couples, via the exchange of four $t$-channel gluons, to the proton target. Calculating the contribution of these diagrams to the $\gamma^* p$ cross section at small~$x$, and isolating the twist-4 contribution one arrives at the conclusion that the contributions to longitudinal polarized photon has the opposite sign compared to the transversely polarized photon. This implies the possibility that, in $F_2= F_L + F_T$ which sums over transverse and longitudinal photons, there is a (partial) cancellation of twist four corrections, whereas the twist-4 corrections to $F_L$ or $F_T$ are larger than the corresponding corrections to $F_2$.

In order to decide whether the HERA data support this possibility, in ~\cite{Bartels:2000hv} 
the saturation model of Golec-Biernat and W\"{u}sthoff ~\cite{GolecBiernat:1998js,GolecBiernat:1999qd} which successfully describes the HERA data has been used to obtain a quantitative estimate of twist-4 and even higher twist corrections. This mathematically fairly simple model has four parameters 
which are fixed by adjusting the model to describe well the HERA $F_2$ data. 
The model leads to the total cross sections that exhibit the geometric scaling~\cite{GS}.
The cross sections obtained in the model may be expanded in powers of 
$\frac{Q^2_0}{Q^2} \left(\frac{1}{x}\right)^{\lambda}$. 
It was natural to identify the first two terms of this expansion as `leading twist-2' and 
`twist-4 correction', respectively.  Despite its simplicity this model is close enough to the 
lowest order QCD calculations and supports the sign structure of the twist-4 correction 
to the transverse and longitudinal cross sections mentioned before. On a quantitative 
level, the twist-4 corrections to $F_L$ and $F_T$ were found to be sizable, whereas in 
$F_2$ they almost cancel. The overall smallness to the $1/Q^2$ corrections to $F_2$ is
consistent with the estimate of ~\cite{MRST,MR}: the analysis in ~\cite{Bartels:2000hv} therefore provides a natural explanation of the suppression of twist-4 corrections to $F_2$,
without demanding that the higher twist contribution is small for $F_L$ or $F_T$. 

This original version of the GBW model did not include any QCD evolution. Therefore,
the connection of this model with evolution of twist four operators in QCD is not possible, and, in particular, the  identification of the first and the second term in the $1/Q^2$ expansion as the `leading' and the `next-to-leading' twist seems somewhat crude.  A more recent version ~\cite{Bartels:2002cj} of the GBW model includes QCD evolution and its description 
of HERA data is slightly better than that of the original model. It is therefore natural to investigate to what extent this improved model exhibits the structure expected for higher twist operators, and then to perform a numerical analysis similar to ~\cite{Bartels:2000hv}. This is the goal the present paper. 

Our numerical analysis shows an interesting pattern of the higher twist effects in the structure functions. The corrections are sizable in $F_L$, for the kinematic range relevant for HERA data at low $Q^2$, $1.8$~GeV$^2 < 10$~GeV$^2$ and small~$x$, the twist-4 corrections are found to reduce the leading twist result by about 20\,--\,50\%. We compare the obtained predictions for $F_L$ to recent HERA data~\cite{FLH1new}, both for the complete saturation model and its leading twist component. For $F_2$ the higher twist effects are found to be surprisingly small, at a few percent level down to $Q^2=1$~GeV$^2$.

The paper is organized as follows. The first, longer part in Secs.\ II--VI is devoted to discussion of theoretical issues, and the second part, in Secs.\ VII--IX , to the phenomenological applications. In section II we review simple QCD calculations of  higher twist corrections in momentum space, restricting ourselves to the double 
logarithmic approximation, and in section III we reformulate these results in the QCD dipole 
picture. Next we turn to the saturation model: after a brief review of the simple
GBW model in section IV, we perform a theoretical twist analysis of the QCD improved dipole model in section V--VI. The second part (section VII--IX) contains our numerical analysis and discussions of the results and possible consequences for physics at the LHC. 
Conclusions are given in section X.

\section{The double leading-log approximation in QCD}  
\label{sec:2}

In this section we give a brief overview of higher twist corrections in small-$x$ QCD. We consider the scattering of a virtual photon with transverse or longitudinal polarization on a quark, and we restrict ourselves to the leading behavior at large $Q^2$ and small $x$ (double logarithmic approximation, DLA). In this limit, we can either start from the leading-log $Q^2$ limit and then take the small-$x$ limit or, alternatively, start from the small-$x$ limit and then investigate the large-$Q^2$ approximation. We begin with the latter one, i.e.\ we 
restrict ourselves to those diagrams which have the maximal number of logarithms in $1/x$.

\subsection{Twist 4 corrections}

In the small-$x$ limit, the scattering of a virtual photon on a quark is described by  
the exchange of 2 or more gluons between a closed quark loop and the target quark.
For two $t$-channel gluons, the coupling to the virtual photon is described by the 
well-known photon impact factor, $D_{(2,0);T,L}(Q;\bk,\bq-\bk)$, where the subscripts $T,L$ refer 
to the polarizations of the virtual photon, and $\bk$ and $\bq-\bk$ are the transverse momenta of  
the two $t$-channel gluons. In the deep inelastic limit: $q^2=0$, $k^2 \ll Q^2$ we have 
~\cite{BW,BB}:
\beeq
\begin{array}{lllrlrr}
D_{(2,0);T}(Q;\bk,-\bk) = & A_0 \left( \right.
                      &\frac{4}{3} \frac{k^2}{Q^2} \log \frac{Q^2}{k^2} 
                      & + \frac{14}{9} \frac{k^2}{Q^2}\hspace{0.5cm} 
                      & 
                      &     + \frac{2}{5}(\frac{k^2}{Q^2})^2\hspace{0.5cm}
                      & \left. + {\cal O}((\frac{k^2}{Q^2})^3)  \right) \\  
     
D_{(2,0);L}(Q;\bk,-\bk) = & A_0  \left( \right.
                      &
                      &+\frac{2}{3}  \frac{k^2}{Q^2}\hspace{0.5cm} 
                      & -\frac{4}{15} (\frac{k^2}{Q^2})^2  \log \frac{Q^2}{k^2} 
                      & - \frac{94}{225} (\frac{k^2}{Q^2})^2 \hspace{0.5cm}
                      & \left. + {\cal O}((\frac{k^2}{Q^2})^3)  \right) 
\end{array} 
\label{eq:D2twist}
\eeeq
where $A_0 = \sum e_f^2 \alpha_s \frac{\sqrt{8}}{2\pi}$. The leading power, $\left(k^2/Q^2\right)$, belongs to leading twist, the terms proportional 
to  $\left(k^2/Q^2\right)^2$ to twist four etc. These results, obtained directly from Feynman diagrams in the momentum space, lead to estimates of twist contributions to $F_T$ and $F_L$ consistent with estimates of the saturation model, discussed later, in Sec.~\ref{sec:3} and Sec.~\ref{sec:4}, obtained in the Mellin representation. Important features of $D_{(2,0);T,L}$ are the logarithms and signs of the 
$\left(k^2/Q^2\right)^2$ corrections: whereas for transverse polarization there is no logarithmic enhancement and the power correction is positive, for longitudinal polarization we have a logarithmic enhancement, and compared to the transverse case is has the opposite sign. These simple observations open the possibility that, in $F_L$, the higher twist corrections are large and that, in the structure function $F_2$ which sums over  transverse and longitudinal polarizations of the photon:
\beq
F_2= F_T +F_L,
\eeq   
the total twist-four corrections may be small due to cancellations. Within the model to be
discussed in this paper we will find that, for the HERA data, this is indeed the case.

\begin{figure}
\begin{center}
\epsfig{file=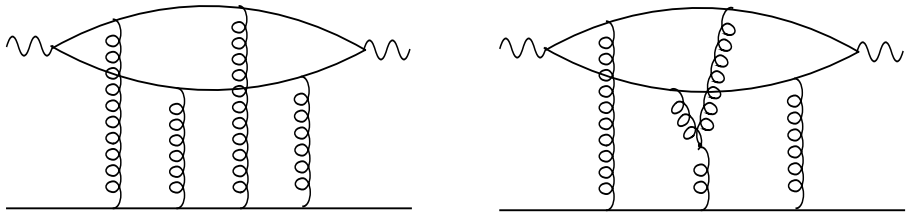,width=12cm,height=3cm}
\end{center}
\caption{
Two diagrams with four $t$-channel gluons.
\label{fig1}}
\end{figure}

The simplest diagram for corrections due to the exchange 
four gluons (order ${\cal O}(g^8)$) is illustrated in Fig.~\ref{fig1}, left. 
In addition to this exchange diagrams, there are diagrams involving the triple gluon vertex,
like e.g.\ the diagram shown in Fig.~\ref{fig1}, right.

An efficient method of calculating the sum of all these contributions in the high energy or small-$x$ limit is the effective action, defined in ~\cite{Lipatov} and further studied in ~\cite{Hentschinski}. As a result 
we have, for all diagrams up to order $g^8$, two classes of contributions:\\
(a) the BFKL ladders~\cite{bfkl1,bfkl2,bfkl3,bfklsum}, expanded up to order $g^8$. This includes NLO corrections to the gluon 
trajectory as well as NLO corrections to the BFKL kernel which, in the leading log approximation, will not be considered;
(b) the exchange of four $t$-channel gluons, where the eikonal couplings to the quark loop at the 
top and to the quark line at the bottom are fully symmetrized, both in color and in their 
momentum structure. The coupling to the quark loop, 
$D_{(4,0);T,L}^{sym}(\bk_1,\bk_2,\bk_3,\bk_4)$, can be expressed in 
terms of $D_2$:    
\begin{align}
{D_{(4,0);T,L}}_{sym}^{a_1a_2a_3a_4}(\bk_1,\bk_2,\bk_3,\bk_4) = -g^2 {d^{a_1a_2a_3a_4}}_{\mathrm{sym}} \nonumber \\ \times \;
\left( \,      D_{(2,0);T,L}(\bk_1+\bk_2+\bk_3,\bk_4) + D_{(2,0);T,L}(\bk_2+\bk_3+\bk_4,\bk_1)  
              \right.\nonumber \\ \left.
             + \, D_{(2,0);T,L}(\bk_3+\bk_4+\bk_1,\bk_2) + D_{(2,0);T,L}(\bk_4+\bk_1+\bk_2,\bk_3)
                 \right.\nonumber \\ \left.
              - \, D_{(2,0);T,L}(\bk_1+\bk_2,+\bk_3 + \bk_4)-D_{(2,0);T,L}(\bk_1+\bk_3, \bk_2+\bk_4)
               - D_{(2,0);T,L}(\bk_1+\bk_4, \bk_2+\bk_3)   \right), 
\label{eq:D4decomposition}
\end{align} 
where $a_i$ and $\bk_i$, $i=1,\ldots,4$ are the color indexes and transverse momenta of the gluons. The color factor has the form:
\beq
{d^{a_1a_2a_3a_4}}_{\mathrm{sym}} =\Big[ tr \left( t^{a_1}t^{a_2}t^{a_3}t^{a_4}\right) 
         + tr \left( t^{a_4}t^{a_3}t^{a_2}t^{a_1}\right) \Big]_{\mathrm{sym}},  
\eeq 
where the subscript `sym' indicates that the color labels are completely symmetrized.

Next let us consider higher order corrections to (a) which, in the leading logarithmic 
approximation, sum up to the BFKL ladders. Using a Mellin representation for the impact 
factors  $D_{(2,0);T,L}(Q;k,-k)$ and a double Mellin representation for the BFKL Green's 
function we write the scattering amplitude in the form:   
\beq
D_{2;T,L}(x;Q^2/Q_0^2) = \int \frac{d\omega}{2\pi i} \int \frac{d\nu}{2\pi i} \left(\frac{1}{x}\right)^{\omega} 
\left( \frac{Q^2}{Q_0^2} \right)^{\nu} D_{(2,0);T,L}(\nu) \frac{1}{\omega - \bar \alpha \chi(\nu,0)}
\eeq
where the integration contours run along the imaginary axis, $\bar \alpha =\frac{N_c \alpha_s}{\pi}$,
the BFKL characteristic function has the form 
\beq
\chi(\nu,0) =2 \psi(1) - \psi (1+\nu) - \psi (-\nu) ,
\eeq   
and $Q_0^2$ denotes the momentum scale at the target end of the BFKL ladder.  
The scattering amplitude $D_2$ can be expanded in powers of $Q^2/Q_0^2$: the terms in this 
expansion are due to the poles of $\chi(\nu)$ at negative integer values: $\nu=-1,-2,...$, and the residues of the poles lead to anomalous dimensions. In particular, the pole near $\nu =-1$ leads to the leading twist behavior  
\beq
\left(Q^2/Q_0^2 \right)^{\gamma (\omega)},\,\,\gamma (\omega) = \frac{N_c \alpha_s}{\pi \omega},
\label{eq:BFKLtwist2}
\eeq
the pole near $\nu =-2$ to the twist four correction  
\beq
\left(Q^2/Q_0^2 \right)^{- 1+ \gamma (\omega)},\,\,\gamma (\omega) = \frac{N_c \alpha_s}{\pi \omega}.
\label{eq:BFKLtwist4}
\ee
The exponents are the anomalous dimension of the higher twist operators contained 
in the BFKL amplitude ~\cite{Lipatov86}. In the notation of~\cite{qBFKL}, they belong to the class of non-quasipartonic operators. When coupled to the impact factors 
$D_{(2,0);T,L}(Q;-k)$, we see from the expansion in (\ref{eq:D2twist}) that, 
for the leading-twist terms, the transverse polarization has a logarithmic enhancement   
compared to the longitudinal polarization. For the twist four correction, the 
situation is reversed. As to the sign structure, the twist four corrections  
to transverse and longitudinal polarizations have opposite signs.    

\begin{figure}
\begin{center}
\epsfig{file=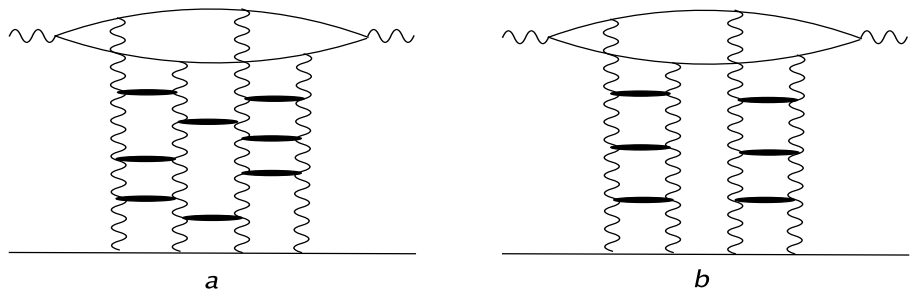,width=12cm,height=4cm}
\end{center}
\caption{
(a) Pairwise interactions between four reggeized gluons; (b) two noninteracting color singlet ladders.
\label{fig2}
}
\end{figure}

Next we turn to higher corrections to class (b). Following the analysis 
of ~\cite{qBFKL} we consider those corrections which 
are obtained by inserting all possible pairwise interactions between the reggeized $t$-channel gluons 
(Fig.~\ref{fig2}a). The large-$Q^2$ behavior comes from the region where the transverse momenta 
of the exchanged gluons are ordered and much smaller than the virtuality of the external photon. 
The leading (in $1/Q^2$) behavior of the four gluon state has been 
discussed in ~\cite{DLA1,DLA2,DLA3}, and we briefly summarize. In the complex $\nu$-plane, 
the leading singularity is a pole at 
\beq
\gamma_{pole}= \frac{4 N_c \alpha_s (1 + \delta)}{\pi \omega}
\label{eq:twist4pole}
\eeq
where $\delta$ is a correction of the order $1/N_c^4$ 
($\delta = 0.778/N_c^4$, and for $N_c=3$, $\delta = 0.0096$).  
In the large-$N_c$ limit the four gluon state reduces to two noninteracting color singlet gluon ladders, leading to a cut in the $\nu$-plane with the branch point located at 
\beq 
\gamma_{cut}= \frac{4 N_c \alpha_s}{\pi \omega}
\label{eq:twist4cut}
\eeq
The pole in (\ref{eq:twist4pole}) at finite $N_c$ can be viewed as `bound state' 
formed by the two color singlet ladders, whereas the cut (\ref{eq:twist4cut}) represents 
the `threshold' of two free ladders. The large-$Q^2$ behavior of this four gluon state is 
described the evolution equations of the twist four gluon operator in the small-$x$ limit, as 
discussed in ~\cite{qBFKL}. In the large-$N_c$ limit, the evolution equations reduce to two 
independent DGLAP ladders.   

In order to apply this discussion to the diagrams of class (b) we notice that the 
twist four contribution of the fermion loop, $D_{(4,0)T,L}$ is easily obtained from 
(\ref{eq:D4decomposition}) and (\ref{eq:D2twist}). For example, the twist four correction
of the transverse polarization is found to be proportional to  
\beq
4 g^2 A_0 \frac{2}{5 } \frac{\bk_1\cdot \bk_3\, \bk_2 \cdot \bk_4 + \bk_1\cdot \bk_2\, \bk_3 \cdot \bk_4 + 
\bk_1\cdot \bk_4\, \bk_2 \cdot \bk_3 }{(Q^2)^2},
\eeq  
and analogous results are found for the longitudinal polarization.
In this way, the four gluon state corrections to transverse and longitudinal 
polarizations follow the same pattern as the twist four piece inside the 
BFKL ladder: compared to transverse polarization, the longitudinal 
polarization has a logarithmic enhancement and comes with the opposite sign.    

Compared to the BFKL-singularity in (\ref{eq:BFKLtwist4}), 
both in (\ref{eq:twist4pole}) and in (\ref{eq:twist4cut})  the coefficients of the 
pole at $\omega=0$ are larger by a factor 4: at small~$x$, this twist-4 correction will therefore  
dominate. This suggests to consider, within a twist expansion in the small-$x$ region, 
as a first set of higher twist corrections these four gluon states, disregarding the higher twist contributions of non-quasipartonic operators.
It is not difficult to generalize this selection to six, eight etc gluon states.
If, in addition, one invokes the large-$N_c$ expansion where the $2n$-gluon 
state is approximated by $n$ noninteracting color singlet ladders, one arrives at the eikonal picture of multi-Pomeron exchange, which underlies the saturation model to be discussed  
further below. We shall see that this model embodies many of the features 
of these $n$-ladder exchanges, in particular the correct $Q^2$-evolution.

\subsection{Higher twists in the Balitsky-Kovchegov equation}

\begin{figure}
\begin{center}\epsfig{file=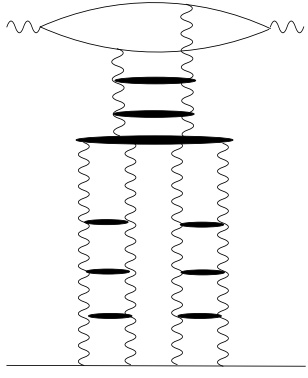,width=4cm,height=5cm}
\end{center}
\caption{
A fan diagram.
\label{fig3}}
\end{figure}

So far we have discussed a selected subclass of QCD diagrams giving rise to twist-4 corrections to the proton structure functions. It should, however, be kept in mind that this selection of higher twist corrections is not in agreement with what one obtains from summing all leading-$\log 1/x$ contributions from the BFKL Pomeron fan diagrams. This summation may be performed using the Balitsky-Kovchegov (BK) equation~\cite{Bal,Kov1,Kov2}.  
To illustrate this, we summarize the results of the complete small-$x$ analysis which can be found in ~\cite{BE,TPV,BKutak}. The sum of all diagrams contributing to the leading logarithmic $1/x$ 
approximation can be organized in two classes.\\
(i) BFKL ladders consisting of reggeized gluons. At the lower end, reggeized gluons can split 
into two or three elementary gluons. In the former case, the color structure of the 
splitting is described by a structure constant $f_{c a_1 a_2}$, in the latter case by the 
product of two structure constants, e.g. $f_{c a_1 d}f_{d a_2 a_3}$.\\   
(ii) BFKL-like ladders where, instead of the reggeized gluons which belong to the adjoint 
color representation, we have Reggeons in the symmetric octet and singlet color representations. In both cases, the trajectory functions are the same as for the reggeized gluon. The corresponding color tensors are listed in  ~\cite{BE}. The sum of these diagrams is symmetric under the exchange of momenta and color indexes. \\  
(iii) Diagrams with a four gluon $t$-channel state. This state is symmetric under the 
exchange of $t$-channel gluons (momenta and color indexes). There is no direct coupling 
of this state to the quark loop at the top. Instead, through the $2 \to 4$ reggeized gluon vertex it couples to a BFKL ladder which then connects with the quark loop (Fig.~\ref{fig3}). Class (i) and (ii) represent the all-order generalizations of (a) and (b), respectively, whereas 
(iii) starts at the order $g^{10}$. In ~\cite{BE}, (i) and (ii) are denoted by $D_4^R$, class (iii) by $D_4^I$. 

As we have already stated, in (iii) the four gluon state that we have discussed before does not couple directly to the quark loop: the coupling goes through a BFKL ladder and a $2 \to 4$ transition vertex. Making use of the large-$Q^2$ results discussed before, we interpret this as a mixing between the non-quasipartonic twist four piece inside the BFKL ladder and the twist four gluon operator. A detailed analysis ~\cite{BKutak}, however, shows that, at the leading logarithmic $\log Q^2$, approximation, this transition kernel between the two twist-4 operators, in the large-$N_c$ limit, vanishes. This also holds for the transition of the twist-6 piece inside the BFKL ladder to the twist-6 piece in the four gluon state. Generalizing this to more than four $t$-channel gluons, one arrives at the conclusion that, in the double leading logarithmic approximation,  the contribution of higher twists given by the BK fan diagrams vanishes, and only propagation of two interacting $t$-channel gluons contributes to the amplitudes. Note however that this result is only valid in the large-$N_c$ limit. 

Therefore, from a theoretical point of view, the twist expansion derived from the  
selection of diagrams with $2n$ $t$-channel gluons in the large-$N_c$ limit 
should be viewed as being different from descriptions based upon the BK equation.  
    
\section{Color dipole picture}
\label{sec:3}

It is important to emphasize that, in the small-$x$ limit, for the class of QCD diagrams 
which we have discussed, 
the scattering amplitude for the elastic scattering of a virtual photon on a quark can be 
cast into the dipole form \cite{Nikolaev:1990ja}:
\beq\label{eq:sigma1}
\sigma^{\gamma^*p}_{T,L}(x,Q^2) \,=\,
\sum_f \int d^2 \rbo \int_0^1\!dz\;
|\Psi_{T,L}^f(z,r,Q^2)|^2 \; \sigmahat (x, \rbo)
\eeq
where $T$ and $L$ denote  the virtual photon polarization: transverse and longitudinal, respectively. The light-cone photon wave function, $\Psi_{T,L}^f$, is modeled by the lowest order  $\gamma^* g \to q \bar q$ scattering amplitudes which give
\begin{eqnarray}
|\Psi_T^f(z,r,Q^2)|^2  &=&  \frac{2N_c\alfaem e_f^2}{4\pi^2}
\left\{\left[z^2+(1-z)^2\right]
\epsilon^2\, K_1^2(\epsilon r) + m_f^2\, K_0^2(\epsilon r)\right\}
\label{eq:overgg}
\\ 
|\Psi_L^f(z,r,Q^2)|^2  &=& 
\frac{8N_c\alfaem e_f^2}{4\pi^2}\,Q^2 z^2(1-z)^2\, K_0^2(\epsilon r)
\label{eq:overgg1} 
\end{eqnarray}
where $K_{0,1}$ are the Bessel--McDonald functions, 
$\epsilon^2={z(1-z)Q^2+m_f^2}$ and $r=|\rbo|$. 
The measured structure functions are related to 
$\sigma^{\gamma^*p}_{T,L}(x,Q^2)$ by the standard formula
\be
F_{T,L}= \frac{Q^2}{4 \pi^2 \alpha_{em}}
\ee
In (\ref{eq:sigma1}), all 
details describing the interaction of the 
quark-antiquark pair with the target quark are contained in the dipole 
cross section, $\sigma(x,\rbo)$.
In particular, the exchange of two non-interacting color singlet gluon ladders provides 
a contribution proportional to the product of two gluon structure functions, 
$(xg(x,C/r^2))^2$. 
 
Important characteristics of the twist expansion follow from the structure of the 
photon wave functions and do not depend upon the details of $\sigma(x,\rbo)$. This is most easily seen by taking 
the Mellin transform of (\ref{eq:sigma1}).      
In general, the Mellin transform
of a function $f(r^2)$ is defined as
\be
\tilde f(s)\equiv{\cal M}_{r^2}\!\left[f(r^2)\right]\!(s)
\,=\int_{0}^{\infty} dr^2\,(r^2)^{s-1} \,f(r^2)
\ee
while the inverse relation reads
\be
f(r^2)\,=\int_C\frac{ds}{2\pi i}\,(r^2)^{-s}\, \tilde f(s)
\ee
where the integration contour $C$ lays in the fundamental strip of the Mellin transform
to be discussed below.

Let us write Eq.~(\ref{eq:sigma1}) in the following form
\beq
\label{eq:sigmah}
\sigma_{T,L}^{\gamma^*p}(x,Q^2)\, = \int_0 ^{\infty} {dr^2 \over r^2} \,
H_{T,L}(r,Q^2) \,\sigmahat(x,r)
\eeq
where 
\beq
H_{T,L}(r,Q^2) \, \equiv \, \pi r^2 \, \sum_f \int_0^1 dz \,
|\Psi_{T,L}^f(z,r,Q^2)|^2\,.
\eeq 
Substituting  the inverse Mellin transform of the dipole cross section,
\be
\sigmahat(x,r)\,=\int_C\frac{ds}{2\pi i}\,(r^2 Q_0^2)^{-s}\,\,\sigmatilde(x,s)
\ee
we find the Mellin representation of the $\gamma^*p$ cross sections, given by the
Parseval formula
\beq
\label{eq:parsival}
\sigma_{T,L}^{\gamma^*p}(x,Q^2)\, = \int\limits_{C_s} {ds \over 2\pi i } \;\sigmatilde(x,s)\,
\tilde H_{T,L}(-s,Q^2/Q_0^2)
\eeq
where $\Htilde_{T,L}(-s,Q^2)$ is the Mellin transform of $H_{T,L}(r,Q^2)$.
The integration contour $C_s$ in the complex $s$-plane is placed in the fundamental strip
in which the integrals defining $\tilde\sigmahat(x,s)$ and $\tilde H_{T,L}(-s,Q^2)$
are convergent.  The strip is determined from the following leading behaviour 
of both functions at small  and large values of $r$ (up to logarithms of $r$): 
\be
H_{T,L}(r,Q^2)=
\left\{ 
\begin{array}{ll}
\mbox{const} & \mbox{for}\;\; r \to 0, \\
{1/r^{2n}} & \mbox{for} \;\; r \to \infty
\end{array}
\right.
\label{eq:hroo}
\ee
with $n=1$ for transverse and $n=2$ for longitudinal polarization.
For the dipole cross section we take, as an example,  
\be\label{eq:sigr0}
\sigmahat(x,r)\,=\left\{ 
\begin{array}{ll}
r^2 & \mbox{for}\;\; r \to 0, \\
\mbox{const} & \mbox{for} \;\; r \to \infty
\end{array}
\right.
\ee
In this case the fundamental strip of $\tilde\sigmahat(x,s)$ is
defined by the condition $-1< \mathrm{Re}\,s  < 0$ while the fundamental strip
of  $\tilde H_{T,L}(s,Q^2)$ is given by $0< \mathrm{Re}\,s  < n$. Taking into account the minus
sign in $\tilde H_{T,L}(-s,Q^2)$, we find that the integration contour ${C_s}$ 
in Eq.~(\ref{eq:parsival}) lays in the strip:  
\be\label{eq:strip}
-1< \mathrm{Re}\,s < 0\,.
\ee
It can be chosen parallel to the imaginary axis, for example, $s=-1/2+i\nu$ with real $\nu$. 

For the Mellin transform of $H_{T,L}(r,Q^2)$ we restrict ourselves to massless quarks, $m_f=0$. 
In this case, $H_{T,L}$ are functions of only one combined variable, $\rhat=r\,Q$:
\beeq
\label{eq:3a}
H_T(\rhat)\eq A_0\intlim_0^1 dz\,[z^2+(1-z)^2]\,z(1-z)\, \rhat^2K_1^2(\sqrt{z(1-z)}\, \rhat)
\\\nonumber
\\
\label{eq:3b}
H_L(\rhat)\eq 4A_0\intlim_0^1 dz\, z^2(1-z)^2\,  \rhat^2K_0^2(\sqrt{z(1-z)}\, \rhat)\,.
\eeeq
where we introduced $A_0={N_c\alpha_{em}\ebar}/{(2\pi)}$ and $\ebar=\sum_f e_f^2$.
In Appendix \ref{app:a} we found the Mellin transforms of these functions in the form
\be\label{eq:3c}
\Htilde_{T,L}(s,Q^2)\,=\left(\frac{Q^2}{4}\right)^{-s}\Htilde_{T,L}(s)
\ee
with  $\Htilde_{T,L}(s)$ given by Eqs.~(\ref{eq:a7}) and (\ref{eq:a7l}):
\be\label{eq:a7T}
\Htilde_{T}(s)\,=\,\frac{A_0\pi}{8}\,
\frac{\Gamma(2+s)\,\Gamma(1+s)\,\Gamma(s)\,\Gamma(1-s)\,\Gamma(3-s)}
{\Gamma(3/2+s)\,\Gamma(2-s)\,\Gamma(5/2-s)}\,.
\ee
and  
\be
\label{eq:a7L}
\Htilde_{L}(s)\,=\,\frac{A_0\pi}{4}\,
\frac{(\Gamma(1+s))^3\,\Gamma(2-s)}{\Gamma(3/2+s)\,\Gamma(5/2-s)}.
\ee
Both functions have simple or multiple  poles for negative and positive real values of $s$. 

Substituting these results into Eq.~(\ref{eq:parsival}) we obtain
\be\label{eq:GBWtwist}
\sigma_{L,T}^{\gamma^*p}(x,Q^2)\, =\, \int\limits_{C_s} {ds \over 2\pi i } 
\left(\frac{Q_{0}^2}{Q^2}\right)^{-s} \tilde \sigma(x,s) \,\,
\Htilde_{T,L}(-s)
\ee
with  the contour $C_s$ in the fundamental strip (\ref{eq:strip}). The twist expansion 
is obtained by closing the $s$-contour to the left. The functions $\Htilde_{T,L}(s)$ have single 
poles to the right of $C$ at positive integers, except for the regular points at $s=2$ for 
transverse and $s=1$ for longitudinal polarizations.
Thus, both functions have the following Laurent expansion around each singular point $s=n$:
\be\label{eq:sigHTL}
\Htilde_{T,L}(s)\,=\,\frac{a_{T,L}^{(n)}}{s-n}\,+\,b_{T,L}^{(n)}\,+\,{\cal O}\left(s-n\right)
\ee
with 
\beeq
a_{T}^{(1)}&=& - \frac{2}{3} A_0,\nonumber\\
 a_{T}^{(2)}&=&0, \,  b_{T}^{(2)} = - \frac{4}{5} A_0 
\eeeq
and 
\beeq
a_{L}^{(1)}&=&0,\, b_{L}^{(1)} = \frac{2}{3} A_0 \nonumber\\
 a_{L}^{(2)}&=&  - \frac{16}{15} A_0.
\eeeq
Assuming, for simplicity, 
that to the left of $C_s$, $\tilde \sigma(x,s)$ has only poles at $s=-1,2,...$ (the more realistic case where 
$\tilde \sigma$ has cuts in the complex $s$-plane will be discussed further below), 
we close the contour to the left and arrive at the twist expansion:
\be\label{eq:GBWtwista}
\sigma_{L,T}^{\gamma^*p}(x,Q^2)\,=\,\sum_{n=1}^\infty\sigma_{T,L}^{(\tau=2n)}(x,Q^2)\,.
\ee
where $\sigma^{(\tau=2n)}\sim 1/Q^{2n}$ (modulo powers of $\log Q^2$). 
With (\ref{eq:a7T}) and (\ref{eq:a7L}) we obtain:
\be
\sigma_{T,L}^{(\tau=2n)}\,=\int\limits_{C_n} {ds \over 2\pi i } 
\left(\frac{Q_{0}^2}{Q^2}\right)^{-s}
\tilde \sigma(x,s) 
\left\{\frac{-a_{T,L}^{(n)}}{s+n}\,+\,b_{T,L}^{(n)}+\ldots\right\}
\ee
where the dots stand for terms regular at $s=-n$. In particular,
for the twist-4 corrections we re-discover the previous result from
(\ref{eq:D2twist}):\\
(i) due to the vanishing of $a_{T}^{(2)}$, the longitudinal structure 
function is enhanced,\\
(ii) the leading terms in $F_T$ and $F_L$ come with opposite signs.

For completeness, we also consider the complex half $s$-plane to the right of the contour $C$.
It is well known that the Bessel-McDonald functions $K_{\nu}(x)$ have a convergent expansion 
around $x=0$, whereas for large arguments the expansion in powers of $1/x$ is asymptotic. 
Therefore, writing the functions  $H_{T,L}(\rhat)$ in the form 
\beeq
H_{T}(\rhat)\eq \frac{A_0\pi}{8}
\int\limits_C\frac{ds}{2\pi i}\left(\frac{\rhat^2}{4}\right)^{-s}\,
\frac{\Gamma(2+s)\,\Gamma(1+s)\,\Gamma(s)\,\Gamma(1-s)\,\Gamma(3-s)}
{\Gamma(3/2+s)\,\Gamma(2-s)\,\Gamma(5/2-s)}
\\\nonumber
\\
H_{L}(\rhat)\eq \frac{A_0\pi}{4}
\int\limits_C\frac{ds}{2\pi i}\left(\frac{\rhat^2}{4}\right)^{-s}\,
\frac{(\Gamma(1+s))^3\,\Gamma(2-s)}{\Gamma(3/2+s)\,\Gamma(5/2-s)}\,,
\eeeq 
we conclude that the expansion in powers of $\rhat$ --- which is obtained by closing the 
contour to the left --- is convergent. In contrast, the expansion in powers of $1/\rhat$ --- 
which corresponds to closing the contour to the right and computing residues of the  
poles at positive integers --- leads to a divergent result which form an asymptotic series 
for $H_{T,L}(\rhat)$  when $\rhat^2\to \infty$: 
\be\label{eq:mel19}
H_{T,L}(\rhat)\,\sim\, 
\frac{h_{T,L}^{(1)}}{\rhat^2}+\frac{h_{T,L}^{(2)}}{\rhat^4}+\frac{h_{T,L}^{(3)}}{\rhat^6}\,
+\ldots,
\ee
where the coefficients $h_{T,L}^{(n)}\sim a_{T,L}^{(n)}$ are equal to:
\beeq\label{eq:mel20}
h_{T}^{(1)}\eq 
{\textstyle {8\over 3}}A_0\,,~~~~~~~~~~~h_{T}^{(2)}=0\,,~~~~~~~~~~~~~~
h_{T}^{(3)}={\textstyle {3072\over 35}}A_0
\\\nonumber
\\
h_{L}^{(1)}\eq 0\,,
~~~~~~~~~~~~~~h_{L}^{(2)}={\textstyle {256\over 15}}A_0\,,
~~~~~~~h_{L}^{(3)}={\textstyle {9216 \over35}}A_0\,.
\eeeq
This asymptotic expansion justifies the large-$r$ behaviour of $H_{T,L}(r,Q^2)$ 
used in the determination of the  fundamental strip (\ref{eq:strip}).
Moreover, returning to (\ref{eq:parsival}) we conclude that, because of the negative sign of the argument of $\tilde H_{T,L}$,
the twist expansion is an asymptotic expansion. 

In conclusion, the opposite sign structure as well as the relative enhancement of the twist-4 
corrections to $F_L$ are general features of the small-$x$ limit in QCD, and they provide 
the possibility that the total twist-4 correction to $F_2$ may become small. 
In the following we choose, for a quantitative estimate, a particular model, the QCD improved dipole model.

\section{The Model}
\label{sec:4}

We aim for the construction of the twist expansion of the proton structure functions
$F_T$ and $F_L$ at small values of the Bjorken variable $x$.
The starting point for our following analysis is the 
GBW saturation model~\cite{GolecBiernat:1998js} and its QCD improved version
which incorporates the leading logarithmic DGLAP evolution \cite{Bartels:2002cj}.

The standard formula  defining the total 
cross section for the scattering of a virtual photon
$\gamma^*_{T,L}(Q^2)$ on a proton $p$ at small value of the Bjorken variable $x$ has already been 
written down in (\ref{eq:sigma1}).
The function $\sigmahat(x,\rbo)$ in Eq.~(\ref{eq:sigma1}) is the color dipole cross section,
describing the interaction of the $q\qbar$ pair with the proton. 
In the original GBW formulation \cite{GolecBiernat:1998js} it depends on the dipole size $r$ and the Bjorken variable $x$, and takes the  following form
\beq\label{eq:sigmagbw}
\sigmahat (x,r)\, =\, \sigma_0 \left\{ 1-\exp\left( -{r^2Q_{\mathrm{sat}}^2(x)/4}\right)\right\}
\eeq
where $Q^2_{\mathrm{sat}}$ is a saturation scale which depends on $x$.
After incorporating the DGLAP evolution for small dipole sizes
the dipole cross section is modeled in \cite{Bartels:2002cj}  as
\beq
\label{eq:sigmad}
\sigmahat (x,r) \,=\,
\sigma_0 \left\{
1-\exp\left(-\Omega(x,r^2)\right)
\right\}
\eeq
where the opacity
\beq\label{eq:sigmad1}
\Omega(x,r^2)\, =\,
{\pi^2r^2\, \alpha_s(\mu^2)\, g(x,\mu^2)\over 3\sigma_0}\,,
\eeq
and $g(x,\mu^2)\equiv xG(x,\mu^2)$ is the gluon distribution (multiplied by $x$) which obeys the DGLAP evolution equation (\ref{eq:b1}) from Appendix \ref{app:b}. The evolution scale $\mu^2$ was originally assumed  to depend on the dipole size in the following way: 
\be\label{eq:sigmad2}
\mu^2\,=\,C/r^2+\mu_0^2\,.
\ee 
Both models of the dipole cross section are eikonal and follow the 
Glauber-Mueller formulae. For the remainder of this section, we restrict ourselves to the 
original GBW model. 

Following our discussion of the previous section, we need the Mellin transform of the 
dipole cross section. In the case of the GBW parameterization (\ref{eq:sigmagbw}), we find
\beeq\nonumber
\label{eq:melGBW}
\sigmatilde(x,s)\eq
\sigma_0\int_0 ^{\infty} \, {dr^2}\, (r^2)^{s-1} 
\left\{1-\exp\left( -{r^2Q_{\mathrm{sat}}^2(x)/4}\right)\right\}
\\\nonumber
\\
\eq
-\sigma_0\left(\frac{Q_{\mathrm{sat}}^2}{4}\right)^{-s}\Gamma(s)\,.
\eeeq
Substituting this result, together with relation (\ref{eq:3c}), into Eq.~(\ref{eq:parsival}), 
we obtain
\be
\label{eq:GBWtwist2}
\sigma_{L,T}^{\gamma^*p}(x,Q^2)\, =\, \sigma_0\int\limits_{C_s} {ds \over 2\pi i } 
\left(\frac{Q_{\mathrm{sat}}^2}{Q^2}\right)^{-s} \left\{-\Gamma(s)\right\}
\Htilde_{T,L}(-s)
\ee
with  the contour $C_s$ in the fundamental strip (\ref{eq:strip}). We see that 
the poles to the left of $C_s$ at negative integers lead to the twist expansion:
\be\label{eq:GBWtwista2}
\sigma_{L,T}^{\gamma^*p}(x,Q^2)\,=\,\sum_{n=1}^\infty\sigma_{T,L}^{(\tau=2n)}(x,Q^2)\,.
\ee
where $\sigma^{(\tau=2n)}\sim 1/Q^{2n}$.
Singularities come from the single poles of the Euler gamma function $\Gamma(s)$ and from the  poles in $\tilde H_{T,L}(-s)$. 
%
In particular, encircling the pole at $s=-n$ by a small counter-clockwise oriented contour $C_n$, and expanding both functions around  this point, we obtain
\be
\sigma_{T,L}^{(\tau=2n)}\,=\sigma_0\int\limits_{C_n} {ds \over 2\pi i } 
\left(\frac{Q_{\mathrm{sat}}^2}{Q^2}\right)^{-s}
\left\{\frac{\gamma^{(n)}_1}{s+n}\,+\,\gamma^{(n)}_0+\ldots\right\}
\left\{\frac{-a_{T,L}^{(n)}}{s+n}\,+\,b_{T,L}^{(n)}+\ldots\right\}
\ee
where the dots denote terms regular at $s=-n$. 
The result of the integration is indeed proportional to $1/Q^{2n}$ with the logarithmic enhancement
coming from the double poles
\be
\sigma_{T,L}^{(\tau=2n)}=\,\sigma_0\,\frac{Q_{\rm sat}^{2n}}{Q^{2n}}\left\{ 
-\gamma^{(n)}_1a_{T,L}^{(n)}\,\log(Q^2/Q_{\mathrm{sat}}^2)\,+\,
\left(\gamma^{(n)}_1b_{T,L}^{(n)}-\gamma^{(n)}_0a_{T,L}^{(n)}\right)
\right\}.
\ee
In particular, we find \cite{Bartels:2000hv} -- for twist-2:
\beeq\label{eq:10a}
\sigma_T^{(\tau=2)}\eq \frac{\alpha_{em}\sigma_0}{\pi}\left<e^2\right>
\frac{Q_{\mathrm{sat}}^2}{Q^2}\left\{\log(Q^2/Q_{\mathrm{sat}}^2)+\gamma_E+1/6\right\}
\\\nonumber
\\\label{eq:10b}
\sigma_L^{(\tau=2)}\eq \frac{\alpha_{em}\sigma_0}{\pi}\left<e^2\right>\frac{Q_{\mathrm{sat}}^2}{Q^2}
\eeeq 
and for twist-4:
\beeq\label{eq:11a}
\sigma_T^{(\tau=4)}\eq \frac{3}{5}\frac{\alpha_{em}\sigma_0}{\pi}\left<e^2\right>\frac{Q_{\mathrm{sat}}^4}{Q^4}
\\\nonumber
\\\label{eq:11b}
\sigma_L^{(\tau=4)}\eq -\frac{4}{5}\frac{\alpha_{em}\sigma_0}{\pi}\left<e^2\right>\frac{Q_{\mathrm{sat}}^4}{Q^4}
\left\{\log(Q^2/Q_{\mathrm{sat}}^2)+\gamma_E+1/15\right\}.
\eeeq 
Notice the negative sign of $\sigma_L^{(\tau=4)}$ and
the lack of logarithm in $\sigma_L^{(\tau=2)}$ and $\sigma_T^{(\tau=4)}$ due to
the singularity structure (\ref{eq:sigHTL}) with $a_{L}^{(1)}=0$ and $a_{T}^{(2)}=0$.

\section{Singularity structure of the DGLAP improved model}
\label{sec:5}

In the DGLAP improved saturation model, the $r$-dependence of the dipole cross section, given by Eq.~(\ref{eq:sigmad}), is rather involved and its exact Mellin transform is not known. However, it is still possible to extract the information 
about the Mellin transform necessary to carry out the twist analysis.
For this purpose it is convenient to use a slightly modified definition of the scale $\mu^2$
in Eq.~(\ref{eq:sigmad}):
\beq\label{eq:newsatscale}
\mu^2 \; = \; \left\{   
\begin{array}{ll} 
C/r^2 \;\;\; & \mbox{ for }\;\; r^2 < C/\mu_0^2,
\\
\mu_0^2 \;\;\; & \mbox{ for }\;\; r^2 \ge C/\mu_0^2\, .
\end{array} 
\right. 
\eeq
Such a modification preserves all the desired features of the original model and
allows to separate the $r^2$-integration range 
of the Mellin transform of the dipole cross section
$\sigmatilde(x,s)$ into two regions: the {\em perturbative} one, defined by the condition $r^2 <C/ \mu_0^2$, in which the gluon density and strong coupling constant are given by  one-loop expressions with the scale $\mu^2 = C/r^2$, and 
the {\em soft} region, defined by the condition $r^2 \ge C/\mu_0^2$, where the
scale is frozen at $\mu^2 = \mu_0^2$: 
Thus
\be\label{eq:dipscsum}
\sigmatilde(x,s)\,=\,\sigmatilde_{\mathrm{pert}}(x,s)+\sigmatilde_{\mathrm{soft}}(x,s)\,.
\ee

In the  soft region the dipole cross section takes the
form of the  GBW saturation model (\ref{eq:sigmagbw}) with  the saturation scale
\be
Q_{\mathrm{sat}}^2(x)=  \frac{4 \pi^2 \alpha_s(\mu_0^2)\, g(x,\mu_0^2)}{3 \sigma_0}\,.
\ee
The contribution from this region to the Mellin transform is given by
\beeq\nonumber
\label{eq:softcont}
\sigmatilde_{\mathrm{soft}}(x,s) \eq
\int_{C/\mu_0^2} ^{\infty} dr^2 \, (r^2)^{s-1} 
\left\{1-\exp\left( -{r^2Q_{\mathrm{sat}}^2(x)/4}\right)\right\}
\\\nonumber
\\
\eq -\sigma_0\left(\frac{Q_{\mathrm{sat}}^2}{4}\right)^{-s}\left\{\frac{a^s}{s}+\Gamma(s,a)\right\}
\eeeq
where $a={CQ_{\mathrm{sat}}^2}/{(4 \mu_0^2)}$ and $\Gamma(s,a)$ is the incomplete gamma function which
has no singularities in the complex  $s$-plane. The soft part has 
only a single pole at $s=0$ which does not contribute to the twist expansion.

\begin{figure}[t]
\begin{center}
\includegraphics[width=6cm]{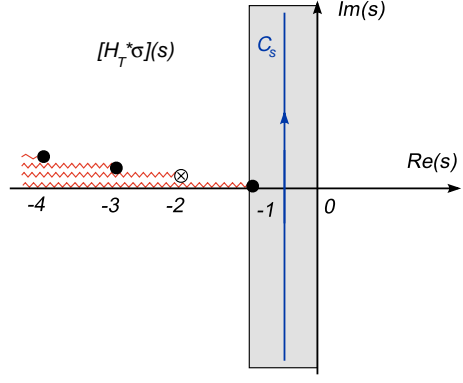}
\end{center}
\caption{ 
The singularity structure of $\sigmatilde(x,s)\Htilde_T(-s)$
in the complex $s$-plane to the left of the Mellin integration contour ${\cal C}_s$
in the fundamental strip shown as the gray band.
The zigzag lines indicate the cuts  coming from powers of the Mellin transform of $\alpha_s(\mu^2)g(x,\mu^2)$ while the full circles are the poles of $\tilde H_T(-s)$ that coincide with  branch points of  $\sigmatilde(x,s)$.  
The crossed circle at $s=-2$ is the branch point that is not accompanied by the pole  of $\tilde H_T(-s)$. 
\label{fig:mel0}
}
\end{figure}

The contribution from the perturbative region may be computed term by term from the expansion
\beeq\nonumber
\label{eq:sigmaexp}
{\sigma_{\mathrm{pert}} (x,r)}\eq
\sum_{n=1} ^{\infty}
{\sigma_{\mathrm{pert}}^{(n)}(x,r)}
\\\nonumber
\\
\eq\sum_{n=1} ^{\infty}
\sigma_0 \,
{(-1)^{n+1}\over n!}\,\, 
\Omega^n_{\mathrm {pert}}(x,r^2)
\eeeq
and the perturbative part of the opacity reads
\beq
\Omega_{\mathrm{pert}}(x,r^2) = {\pi^2 \over 3\sigma_0}\,r^2 
\alpha_s(C/r^2)\, g(x,C/r^2)\, \Theta\left(C/r^2 -\mu_0^2\right).
\eeq
The Mellin transform $\sigmatilde_{\mathrm{pert}}(x,s)$ exists due to the theta distribution and is given by the sum of the Mellin transforms of the subsequent terms in the expansion (\ref{eq:sigmaexp})
\be
{\sigmatilde_{\mathrm{pert}}^{(n)}(x,s)}=\sigma_0 \,
{(-1)^{n+1}\over n!}\,\, 
\widetilde{\Omega^n}_{\mathrm{\!\!\!\! pert}}(x,s)
\ee
Each term contributes a cut singularity in the $s$-plane extending to the left from the branch point at negative integers, 
see Fig.~\ref{fig:mel0}. The positions of the branch points are determined by the corresponding power of $r^2$ since the powers of $\alpha_s(\mu^2)g(x,\mu^2)$
do not introduce any additional shift.

For example,   we compute first the Mellin transform
\be
{\cal M}_{r^2}\!\left[\alpha_s({C/ r^2})g(x,{C/r^2})\Theta\!\left({C/r^2} -\mu_0^2\right)\right](s) 
\,=\left(\frac{C}{\Lambda^2}\right)^{s}\widetilde{\alpha_s g}(x,s)
\label{eq:mu2r}
\ee
where $\widetilde{\alpha_s g}$ denotes the Mellin transform 
with respect to the scale $\mu^2$, defined as  
\be\label{eq:mellinmu}
\widetilde{\alpha_s g}(x,s)\,=\int_{\mu^2_0}^\infty
\frac{d\mu^2}{\mu^2}\left(\frac{\mu^2}{\Lambda^2}\right)^{-s}\alpha_s(\mu^2)g(x,\mu^2)\,.
\ee
In such a case the inverse relation reads
\be\label{eq:mellinmuinv}
\alpha_s(\mu^2)g(x,\mu^2)\,=\int_C\frac{ds}{2\pi i}\left(\frac{\mu^2}{\Lambda^2}\right)^s
\widetilde{\alpha_s g}(x,s)\,.
\ee
where the integration contour lays to the right of the right-most singularity.
Using the property of the Mellin transform: ${\cal M}_t[t^nf(t)](s)\,=\,{\cal M}_t[f(t)](s+n)$, 
we find
\beq
\label{eq:omega1}
{\cal M}_{r^2}[\Omega_{\mathrm{pert}}(x,r^2)](s) 
\,=\,
{\pi^2 \over 3\sigma_0}\; \left(\frac{C}{\Lambda^2}\right)^{s+1} \;\widetilde{\alpha_s g}(x,s+1)
\eeq  
and after substituting  solution (\ref{eq:asgluon1}), we  obtain
\beq
\label{eq:omega1a}
{\tilde\Omega_{\rm pert}}(x,s) 
\,=\,{\pi^2 \over 3\sigma_0}\;  \left(\frac{C}{\Lambda^2}\right)^{s+1}\int {d\omega \over 2\pi i}\, x^{-\omega}\;
\frac{2\pi\,\tilde g_0(\omega)}{\tilde P_{gg}(\omega)}\,\,
(s+1)^{-{b_0\over {2\pi}} \tilde P_{gg}(\omega)} \,.
\eeq  
The logarithmic cut singularity along the negative real axis with the branch point at $s=-1$ is obvious from this solution. 
The Mellin transform of $\Omega^2_{\mathrm{pert}}(x,r^2)$ is given by
\beeq
\label{eq:omega2}
{\cal M}_{r^2}[\Omega^2_{\mathrm{pert}}(x,r^2)](s) \eq
\left( {\pi^2 \over 3\sigma_0 } \right)^2 \; 
\left(\frac{C}{\Lambda^2}\right)^{s+2}\widetilde{(\alpha_s g)^2}(x,s+2)
\\\nonumber
\\
\eq
\left( {\pi^2 \over 3\sigma_0 } \right)^2 \; 
\left(\frac{C}{\Lambda^2}\right)^{s+2}\int {ds' \over 2\pi i}\;
\widetilde{\alpha_s g}(x,s')\; \widetilde{\alpha_s g}(x,s+2-s')\,.
\eeeq
where $\widetilde{(\alpha_s g)^2}$ is the Mellin transform (\ref{eq:mellinmu}) of the product
$[\alpha_s(\mu^2)g(x,\mu^2)]^2$, and the  Mellin convolution theorem was used in the last equality.
It can be shown explicitly that expression (\ref{eq:omega2}) has a  cut singularity along the real axis for $-\infty<s<-2$ with the branch point at $s=-2$.

In general, we have
\beq
\label{eq:omegan}
{\cal M}_{r^2}[\Omega^n_{\mathrm{pert}}(x,r^2)](s) \; = \;
\left( {\pi^2 \over 3\sigma_0 } \right)^n \; 
\left(\frac{C}{\Lambda^2}\right)^{s+n} 
\widetilde{(\alpha_s g)^n}(x,s+n)
\eeq
with  the logarithmic cut along the negative real axis starting at the branch point at $s=-n$,
see Fig.~\ref{fig:mel0}. In summary, the singularity structure
of the Mellin transform (\ref{eq:dipscsum}), relevant for the twist expansion,
is determined by the perturbative part only.

\section{Twist decomposition in the DGLAP improved model}
\label{sec:6}

At each twist the saturation model incorporates a few distinct 
contributions that have a clear interpretation within perturbative QCD.
The contributions may be classified using the
singularity structure of the product $\sigmatilde(x,s) \Htilde_{T,L}(-s,Q^2)\,$
in the Mellin plane.

\subsection{Twist-2 contributions}

\begin{figure}[t]
\begin{center}
\includegraphics[width=6cm]{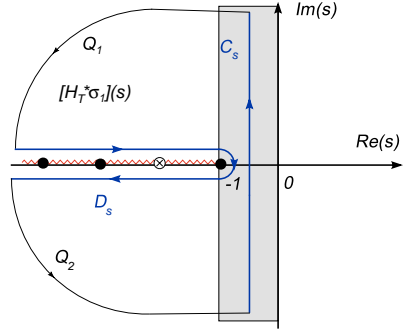} 
\end{center}
\caption{
The discontinuity structure of $\tilde\sigma^{(1)}_{\rm pert}(x,s)\tilde H_T(-s)$ and the
integration contours in the complex $s$-plane with the pieces: ${\cal Q}_1,\,
{\cal Q}_2$ and $D_s$. The meaning of all symbols is as in Fig.~\ref{fig:mel0}
\label{fig:melt2}}
\end{figure}

Starting from the twist-2 analysis, we close the contour ${\cal C}_s$ of the Mellin  integration in Eq.~(\ref{eq:parsival}) with two large quarter-circles ${\cal Q}_1$ and ${\cal Q}_2$ and a 
contour ${\cal D}_s$ enveloping the complex cut of $\sigmatilde(x,s)$ with the branch point at $s=-1$, see Fig.\ \ref{fig:melt2}. Then, we decompose $\sigmatilde(x,s)$ into a 
part which singular at $s=-1$, given by $\sigmatilde^{(1)}_{\rm pert}(x,s)=\sigma_0\tilde\Omega_{\rm pert}(x,s)$,  and a part which is regular at this point, $\sigmatilde_{\rm reg}^{(s=-1)}(x,s)$.
The latter part consists both the soft contribution (\ref{eq:softcont}) and the
contributions from multiple exchanges with cuts starting from $s=-2$.
Thus,  using  expansion (\ref{eq:sigHTL}) for  $\Htilde_{T,L}(-s)$ with $n=1$, we obtain
the  twist-2 part in the form
\be\label{eq:tw21}
\sigma_{T,L}^{(\tau=2)}\,=\int\limits_{-D_s}\frac{ds}{2\pi i}
\left(\frac{Q^2}{4}\right)^s
\left\{\sigma_0\,\tilde\Omega_{\rm pert}(x,s) \,+\,\sigmatilde_{\rm reg}^{(s=-1)}(x,s)
\right\}
\Big\{
\frac{-a_{T,L}^{(1)}}{s+1}\,+\,b_{T,L}^{(1)}\,+\,{\cal O}(s+1)
\Big\}
\ee
where the integration contour is reversed with respect to the contour $D_s$ shown in Fig.~\ref{fig:melt2}, and  the Laurent expansion coefficients
\be
a_{T}^{(1)}=-{\textstyle\frac{2}{3}}A_0\,,~~~~~~~~~~
a_{L}^{(1)}=0\,,~~~~~~~~~~
b_{T}^{(1)}=\left(\textstyle\frac{4}{3}\gamma_E-\textstyle\frac{5}{9}\right)A_0\,,~~~~~~~~~~
b_{L}^{(1)}={\textstyle\frac{2}{3}}A_0\,.
\ee
with $A_0={N_c\alpha_{em}\ebar}/{(2\pi)}$.

Let us compute the twist-2 contribution for transverse photons coming from the most singular
part of the Mellin integrand:
\be\label{eq:tw222}
\Delta_a\sigma_{T}^{(\tau=2)}\,=\,\sigma_0\!\int\limits_{-D_s}\frac{ds}{2\pi i}
\left(\frac{Q^2}{4}\right)^s\,
\Omegatilde_{\rm pert}(x,s) \,
\Big\{
\frac{-a_{T}^{(1)}}{s+1}\Big\}\,.
\ee
The analogous longitudinal contribution vanishes since $a_{L}^{(1)}=0$. 
Using relation (\ref{eq:omega1}), we find
\be\label{eq:tw223}
\Delta_a\sigma_{T}^{(\tau=2)}=\,-\frac{4\pi^2a_{T}^{(1)}}{3\,Q^2}\!
\int\limits_{-D_s}\frac{ds}{2\pi i}
\left(\frac{CQ^2}{4\Lambda^2}\right)^{s+1}\,
\frac{\widetilde{\alpha_s g}(x,s+1)}{s+1}\,.
\ee
The contour integration can be computed directly after substituting Eq.~(\ref{eq:omega1a})
or, alternatively, one can realize that the integral in (\ref{eq:tw223})
is the inverse Mellin transform (\ref{eq:mellinmuinv}) at the scale
$\mu^2=CQ^2/4$ of the following function
\be
S_T^{(2)}(x,\mu^2)\,=\int_{\mu_0^2}^{\mu^2}
\frac{d\mu^{\prime 2}}{\mu^{\prime 2}}\,\,\alpha_s(\mu^{\prime 2})g(x,\mu^{\prime 2})
\,+\,S_T^{(2)}(x,\mu_0^2)\,,
\ee
where the reminder $S_T^{(2)}(x,\mu_0^2)$ depends only on the gluon distribution at an initial scale $\mu^2_0$. It can be computed using the DGLAP equation (\ref{eq:b12})
\be
S_T^{(2)}(x,\mu^2_0)=\int\limits_{C_{\omega}} {d\omega \over 2\pi i}\,\, x^{-\omega}\;\frac{2\pi\,\tilde g(\omega,\mu_0^2)}{\tilde P_{gg}(\omega)}\,.
\ee
Thus, we finally obtain 
\be\label{eq:asigma2}
\Delta_a\sigma_{T}^{(\tau=2)}=\,-\frac{4\pi^2a_{T}^{(1)}}{3\,Q^2}\left\{\int_{\mu_0^2}^{CQ^2/4}
\frac{d\mu^{\prime 2}}{\mu^{\prime 2}}\,\,\alpha_s(\mu^{\prime 2})\,g(x,\mu^{\prime 2})
\,+\,S_T^{(2)}(x,\mu_0^2)\right\}.
\ee
The leading logarithmic term in Eq.~(\ref{eq:asigma2}) coincides with the standard DGLAP expression for $\sigma^{\gamma^* p}_T$ obtained assuming 
that the see quarks come from the gluon splitting in the last step
of the evolution.

The higher orders in the Laurent expansion of $\tilde H_{T,L}(-s)$ in Eq.~(\ref{eq:tw21}),
beyond the singular term, correspond to higher order terms in the
perturbative expansion of the twist-2 contribution.
The next-to-leading order (NLO) contribution originate from the constant 
term  $b^{(1)}_{T,L}$. The  obtained expression is of the form (\ref{eq:tw223}) without $(s+1)$ in the denominator. Thus, we immediately obtain 
\be
\left.
\Delta_b\,\sigma_{T,L}^{(\tau=2)}\right|_{\rm NLO}\,=\,\frac{4\pi^2 b_{T,L}^{(1)}}{3\,Q^2}\,\,
\alpha_s(CQ^2/4)\,g(x,CQ^2/4)\,,
\ee
which for the transverse polarization carries one power of the large logarithm $\log Q^2$ less than the leading term in $\Delta_a\sigma^{(\tau=2)} _{T}$. 
Notice that, as expected, for the longitudinal polarization the first non-vanishing
twist-2 contribution enters at the NLO level.
A similar procedure could  also be applied to higher terms of the Laurent 
series, giving contributions with successively decreasing power of $\log Q^2$. 
Obviously, these higher order terms do not exhaust all the
higher order QCD effects. They are parts of the 
QCD corrections to the twist-$2$ amplitude which come from inclusion
of the quark transverse momentum in the quark box beyond the collinear limit

So far we have dealt with the singular part of $\tilde \sigma(x,s)$ at $s=-1$,
generated by the first term in the perturbative part  of the Glauber-Mueller series (\ref{eq:sigmaexp}) proportional to the gluon distribution $g(x,\mu^2)$. 
The remaining terms of this series as well as the soft part
$\sigmatilde_{\rm soft }(x,s)$ are regular at  $s=-1$. However, they contribute to twist-2  through the pole of $H_{T}(-s)$ at this point, giving
\be
\Delta_c\sigma_{T}^{(\tau=2)}\,=\,-\frac{4a_T^{(1)}}{Q^2}\,\,
\sigmatilde_{\rm reg}^{(s=-1)}(x,s=-1)\,.
\ee
The function on the r.h.s is a sum of two pieces:
the soft part, $\sigmatilde_{\rm soft }(x,s=-1)$, and the Mellin transform of the regular part of the perturbative component, $\sigma_{\rm pert}(x,r)-\sigma_{\rm pert}^{(1)}(x,r)$, computed for
$s=-1$. Thus
\be
\sigmatilde_{\rm reg}^{(s=-1)}(x,s=-1)\,=\int_0^{C/\mu_0^2}\frac{dr^2}{r^4}
\left\{\sigma(x,r)-\sigma_{\rm pert}^{(1)}(x,r)\right\}\,+\,\sigmatilde_{\rm soft }(x,s=-1)\,.
\ee 

We summarize by displaying the most leading twist-2 contribution
to the $\gamma^*p$ cross sections,  obtained
in the DGLAP improved saturation model (with $C=4$):
\beeq\label{eq:tw2tll}
\sigma_{T} ^{(\tau=2)}\eq 
\frac{8\pi^2A_0}{9}\,\frac{1}{Q^2}
\int_{\mu_0^2} ^{Q^2} {d\mu'^2 \over \mu'^2}\,\, \alpha_s(\mu'^2)\,g(x,\mu'^2) 
\\\nonumber
\\\label{eq:tw2lll}
\sigma_{L} ^{(\tau=2)} \eq
\frac{8\pi^2 A_0}{9}\,\frac{1}{Q^2}\,\alpha_s(Q^2)\,g(x,Q^2)\,.
\eeeq
Notice the similarity concerning leading logarithms between the twist-2 contributions in the original GBW model, Eqs.~(\ref{eq:10a}) and (\ref{eq:10b}),  and the above formulae.

\subsection{Twist-4 contributions}

\begin{figure}[t]
\begin{center}
\includegraphics[width=6cm]{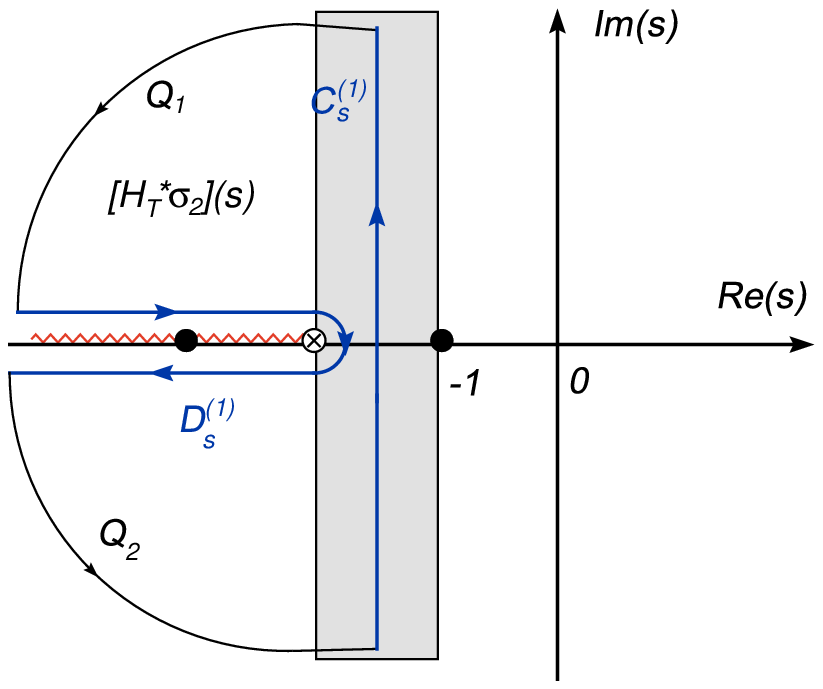} 
\hskip 1cm 
\includegraphics[width=6cm]{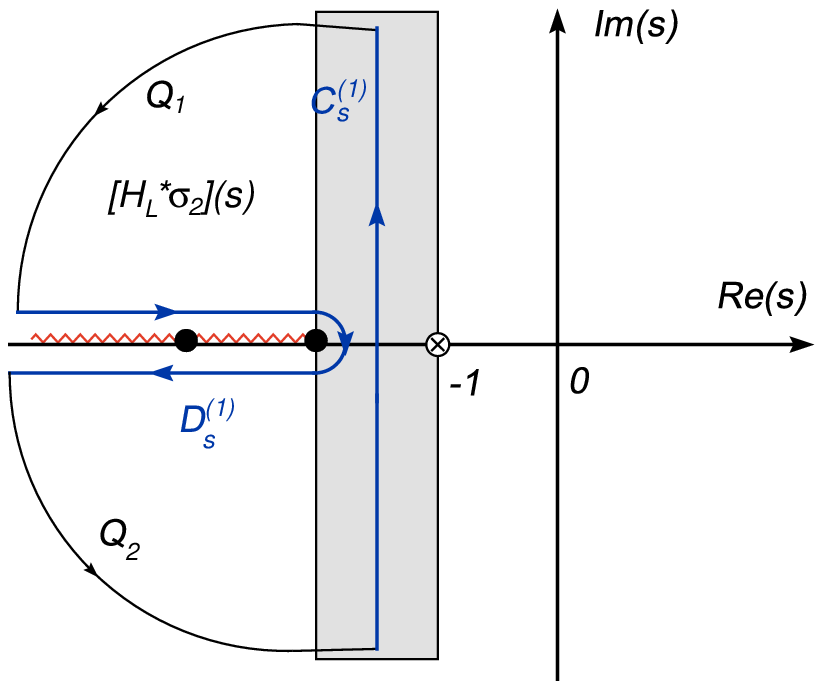} 
\end{center}
\caption{ 
The singularity structure in Mellin plane relevant for twist-4 transverse and longitudinal contributions together with the integration contours. 
The meaning of all symbols is as in Fig.\ \ref{fig:mel0},
but the original contour, ${\cal C}_s$, is replaced by the shifted 
${\cal C}_s ^{(1)}$. The fundamental strip is also shifted.
\label{fig:melt4}}
\end{figure}

The formula for twist-4 is determined by the Mellin transform 
$\sigmatilde^{(2)}_{\rm pert}(x,s)$ of the second term in Eq.~(\ref{eq:sigmaexp}) and the Laurent expansion of $\Htilde_{T,S}(-s)$ around $s=-2$:
\be\label{eq:tw41}
\sigma_{T,L}^{(\tau=4)}\,=\int\limits_{-D_s^{(1)}}\frac{ds}{2\pi i}
\left(\frac{Q^2}{4}\right)^s
\left\{-\frac{\sigma_0}{2}\,\widetilde{\Omega^2}_{\rm \!\!\!pert}(x,s) \,+\,\sigmatilde_{\rm reg}^{(s=-2)}(x,s)
\right\}
\Big\{
\frac{-a_{T,L}^{(2)}}{s+2}\,+\,b_{T,L}^{(2)}\,+\,{\cal O}(s+2)
\Big\}
\ee
where the integration contour envelopes the cut singularity with the branch point at
$s=-2$ in which the function $\sigmatilde_{\rm reg}^{(s=-2)}(x,s)$ is regular, see Fig.~\ref{fig:melt4}. 
The Laurent expansion coefficient are now given  by
\be
a_{T}^{(2)}=0\,,~~~~~~~~~~
a_{L}^{(2)}=-{\textstyle\frac{16}{15}}A_0\,,~~~~~~~~~~
b_{T}^{(2)}=-{\textstyle\frac{4}{5}}A_0\,,~~~~~~~~~~
b_{L}^{(2)}=\left(\textstyle\frac{32}{15}\gamma_E-\textstyle\frac{344}{225}\right)A_0\,.
\ee
The vanishing $a_{T}^{(2)}$ means that the leading logarithmic twist-4 contribution
\be
\Delta_a\sigma^{\tau=4)}_{T}\,=\,\frac{\sigma_0\,a_{T}^{(2)}}{2} \!\!
\int\limits_{-D_s^{(1)}}\frac{ds}{2\pi i}
\left(\frac{Q^2}{4}\right)^s\,
\frac{\widetilde{\Omega^2_{\rm pert}}(x,s)}{s+2}\,,
\ee
vanishes for transverse photons. For the longitudinal polarization it can be found in a similar way as for twist-2, with the following result
\be\label{eq:twLL}
\Delta_a\sigma_{L}^{(\tau=4)}=\,\frac{8\sigma_0 a_L^{(2)}}{Q^4}\left(\frac{\pi^2}{3\sigma_0}\right)^2
\left\{\int_{\mu_0^2}^{CQ^2/4}
\frac{d\mu^{\prime 2}}{\mu^{\prime 2}}\left[\alpha_s(\mu^{\prime 2})\,g(x,\mu^{\prime 2})\right]^2
\,+\,S_L^{(4)}(x,\mu_0^2)\right\}
\ee
where the reminder is of non-perturbative origin and does not depend on $Q^2$,
\be
S_L^{(4)}(x,\mu_0^2)\,=\,
{b_0^2 \over \log\left( {\mu_0 ^2 \over \Lambda^2} \right) }\;
\int {d\omega \over 2\pi i} \, x^{-\omega} \,
\int {d\omega' \over 2\pi i}\; 
{\tilde g_0(\omega') \, \tilde g_0(\omega-\omega')  \over
{b_0 \over 2\pi}\tilde P_{gg}(\omega') 
+ {b_0 \over 2\pi}\tilde P_{gg}(\omega-\omega') -1} \,.
\ee
 
The NLO correction to twist-4 comes from the constant term, $b_{T,L}^{(2)}$, in the Laurent expansion of $\Htilde_{T,L}(-s)$ around $s=-2$. It is straightforward to obtain
\be
\left.
\Delta_b\,\sigma_{T,L}^{(\tau=4)}\right|_{NLO}\,=\,-\frac{8\sigma_0 b_{T,L}^{(2)}}{Q^4}\left(\frac{\pi^2}{3\sigma_0}\right)^2
\left[\alpha_s(CQ^2/4)\,g(x,CQ^2/4)\right]^2\,,
\ee
which in the longitudinal case has one logarithm of $Q^2$ less then the leading contribution 
(\ref{eq:twLL}). As for twist-2, the higher terms in the Laurent expansion
of $\tilde H_{T,L}(-s)$ give rise to yet higher order perturbative corrections.

Multiple scattering effects (with $n\geq 3$) and the soft
contribution are important only for the longitudinal twist-4,  $\sigma^{(\tau=4)}_{L}(x,Q^2)$. They are collected in 
\beq
\Delta_c\sigma^{(\tau=4)} _{L} \, = \,
-\frac{16 a_L ^{(2)}}{Q^4}
\left[ \tilde\sigma_{\mathrm{reg}}^{(s=-2)} (x,s=-2) \, + \, 
\tilde\sigma_{\mathrm{soft}}(x,s=-2) \right]
\eeq
where
\beq
 \tilde\sigma_{\mathrm{reg}}^{(s=-2)}(x,s=-2) \; = \; 
\int_0 ^{C/\mu_0^2} \, {dr^2 \over r^6} 
\left[ \sigma(x,r) - \sigma^{(1)}(x,r) - \sigma^{(2)}(x,r)\right].
\eeq

In summary, the following leading logarithmic structure is found for twist-4
(with $C=4$)
\beeq\label{eq:tw4tll}
\sigma_{T} ^{(\tau=4)} \eq \frac{32\pi^4A_0}{45\,\sigma_0}\,\frac{1}{Q^4}
\left[\alpha_s(Q^2)\,g(x,Q^2)\right]^2
\\\nonumber
\\\label{eq:tw4lll}
\sigma_{L} ^{(\tau=4)} \eq -\frac{128\pi^4A_0}{135\,\sigma_0}\,\frac{1}{Q^4}
\int_{\mu_0^2} ^{Q^2} {d\mu'^2 \over \mu'^2}\left[ \alpha_s(\mu'^2)\,g(x,\mu'^2)\right]^2\,, 
\eeeq
which should be compared to the results obtained in the original GBW saturation model, 
Eqs.~(\ref{eq:11a}) and (\ref{eq:11b}). Notice the similarity in the sign and the leading logarithmic structure.

\subsection{Discussion}

The results (\ref{eq:tw2tll}),(\ref{eq:tw2lll}) and (\ref{eq:tw4tll}),(\ref{eq:tw4lll}) on the leading logarithmic behaviour of the twist-2 and twist-4 contributions are quite general.
For the nucleon structure functions $F_T$ and $F_L$ they imply that
the relative twist-4 correction to $F_T$ is strongly suppressed w.r.t. the 
twist-2 contribution since the subleading twist-4 term in $F_T$ appears
only at the NLO. On the contrary, for $F_L$ the leading twist term
enters only at the NLO and the the twist-4 correction enters 
at the leading order. So, the relative twist-4 effects in $F_L$ are expected to
be enhanced. Note that both in the case of  $F_T$ and $F_L$, the  
twist-4 effects are enhanced w.r.t. the twist-2 contribution by an
additional power of the gluon density $g(x,Q^2)$. 
At sufficiently small $x$, when the gluon density is large, this enhancement 
may compensate the twist-4 suppression w.r.t. the leading twist-2 contribution.

For the structure function $F_2 = F_T + F_L$ we expect small relative 
corrections from the higher twists because of the opposite
sign of the terms proportional to $a^{(2)}_{L}$ and 
$b^{(2)}_{T,L}$. In fact, both  $a^{(2)}_{L}$ and $b^{(2)}_{T,L}$ are 
negative. Thus it follows from (\ref{eq:tw4tll}) and (\ref{eq:tw4lll}) that the resulting 
LO twist-4 contribution to $F_2$ coming from $F_L$ is positive and both 
the dominant (though NLO) term in $F_T$ and the 
NLO correction to $F_L$ are negative. 
This leads to partial cancellation between the twist-4 LO and NLO 
contributions to $F_2$ at moderate $Q^2$, which can be also viewed as a 
partial cancellation between the twist-4 corrections to $F_L$ and $F_T$.

These conclusions about the importance of the higher twist corrections 
are expected to be quite general because they follow directly from the 
twist structure of the quark box and do not depend on the detailed form 
of the twist-4 gluon distribution. In fact, for a generic twist-4 gluon 
density $G_4(x,Q^2)$ (not necessarily proportional to $[g(x,Q^2)]^2$), 
the qualitative pattern of the computed twist-4 corrections emerges.
This happens because independently of the detailed form of gluon
density, the perturbative color dipole scattering cross section 
at twist-4 is proportional to $r^4\,\alpha_s(C/r^2)\, G_4(x,C/r^2)$.
Using a generally valid relation: 
$\int (ds/ 2\pi i) \, x^{-s} \, \tilde f(s)/s = \int^x dx' f(x')$,
one finds
\begin{equation} 
\label{eq:twist4qualit}
\sigma_{T} ^{(\tau=4)}\; \sim  \;
{b^{(2)} _T \over Q^4}\, \alpha^2 _s(Q^2) \, G_4(x,Q^2)\,,
\end{equation}
and
\begin{equation}
\label{eq:twist4qualil}
\sigma_{L} ^{(\tau=4)}\; \sim  \;
-{a^{(2)} _L\over Q^4} \,
\int ^{Q^2} {d\mu'^2 \over \mu'^2}\,\alpha^2_s(\mu'^2)\,
G_4(x,\mu'^2)
\; + \;
{b^{(2)} _L \over Q^4}\, \alpha^2 _s(Q^2) \, G_4(x,Q^2)\, .
\end{equation}
This confirms that for twist-4 the pattern of cancellations in $F_2$ between
$F_L$ and $F_T$ (or between LO and NLO terms) is indeed  universal.

\section{Twist decomposition in the coordinate space}
\label{sec:7}

The preceding analysis was carried out in the Mellin space. This
representation is useful to understand the key features of the twist 
decomposition and match the DGLAP improved saturation model with QCD.
However, in the explicit calculations of the twist composition of 
the $\gamma^* p$ cross sections we find it more convenient to
use the coordinate representation. The main reason is that 
the multiple scattering contributions are represented as multiple 
convolutions in the Mellin space and as simple powers of 
$r^2\alpha_s(\mu^2)g(x,\mu^2)$ in the coordinate space.
Thus, we shall construct an explicit prescription that facilitates the 
twist decomposition
in the coordinate space. The obtained formula (\ref{eq:masterr}) 
is equivalent to its counterpart in the Mellin moment representation
and it  will be used to provide numerical estimates of the twist 
decomposition of the nucleon structure functions.

The singularity structure of the product 
$\,\tilde \sigma(x,s) \, \tilde H_{T,L} (-s,Q^2)\,$ is similar to 
the structure of $\tilde\sigma(x,s)$ except of the branch points of 
$\tilde\sigma(x,s)$ which are strengthened by the poles of $\tilde H_{T,L} (-s,Q^2)$.
In what follows, we shall isolate the contributions of the  singularities
emerging at $s=-1,-2,\ldots$ and link them with 
the twist contributions  $\tau=2,4,\ldots$, respectively 
Let us define two sets of functions,
\beq
\sigma^{(n)}(x,r) = \sigmahat(x,r) \,-\, 
\sigma_0 \, \sum_{k=1} ^n \, {(-1)^k \over k!} \, \Omega^k (x,r^2)
\label{eq:sigsub}
\eeq
and after introducing $\rhat=rQ$
\beq
H_{T,L}^{(n)}(\rhat) = H_{T,L}(\rhat) \,-\, 
\sum_{k=1} ^n \, {h^{(k)}_{T,L} \over (\rhat)^{2k} }
\eeq
where $h^{(k)} _{T,L}$ are the coefficients of the asymptotic expansion  of
$H_{T,L}(\rhat)$, see Eq.~(\ref{eq:mel20}). Additionally, 
$\sigma^{(0)}=\sigmahat$ and $H_{T,L}^{(0)}=H_{T,L}$. 
We see that $\sigma^{(n)}$ describes the contribution to the dipole 
cross section of  $(n+1)$ and more scatterings and 
$H_{T,L}^{(n)}$ gives the contribution to the photon wave function 
from the poles of twist $\tau=2(n+1)$ and higher. The new functions have the
following asymptotics at small and large values of $r$
(modulo logarithms): 
\beq
\sigma^{(n)} (x,r) \;\; \sim \;\;
\left\{ 
\begin{array}{ll}
(r^2)^{n+1} & \mbox{for}\;\; r \to 0
\\ 
(r^2)^n     & \mbox{for} \;\; r \to \infty 
\end{array}
\right.
\label{eq:sigr0n}
\eeq
and
\beq
H^{(n)} _{T,L}(\rhat) \;\; \sim \;\;
\left\{ 
\begin{array}{ll}
1/\rhat^{2n}  & \mbox{for}\;\; r \to 0 \\
 1/\rhat^{2(n+1)}& \mbox{for}\;\; r \to \infty\,. 
\end{array}
\right. 
\label{eq:hroon}
\eeq
Now, it is easy to prove that  $\sigma^{(n)}(x,r)$ and $H^{(n)}_{T,L}(\rhat)$ have 
Mellin transforms, $\tilde \sigma^{(n)}(x,s)$ and $\tilde H^{(n)}_{T,L}(-s,Q^2)$, with the fundamental strip:
\be
-(n+1) < \mathrm{Re}\,s < -n\,.
\ee 
It is moved to the left by $n$~units with respect to the fundamental
strip   given by Eq.~(\ref{eq:strip}).

\begin{figure}[t]
\begin{center}
\includegraphics[width=6cm]{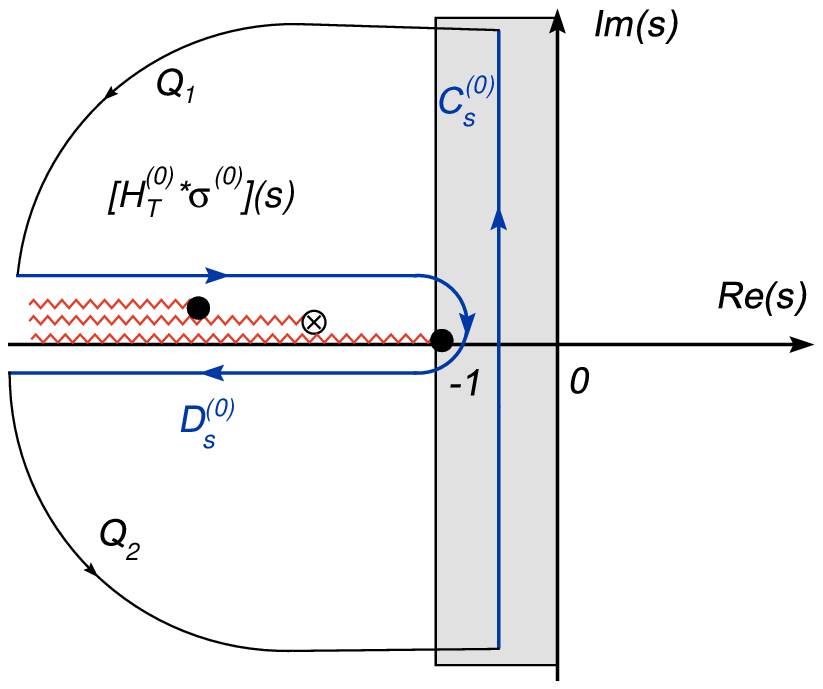}
\hskip 1cm
\includegraphics[width=6cm]{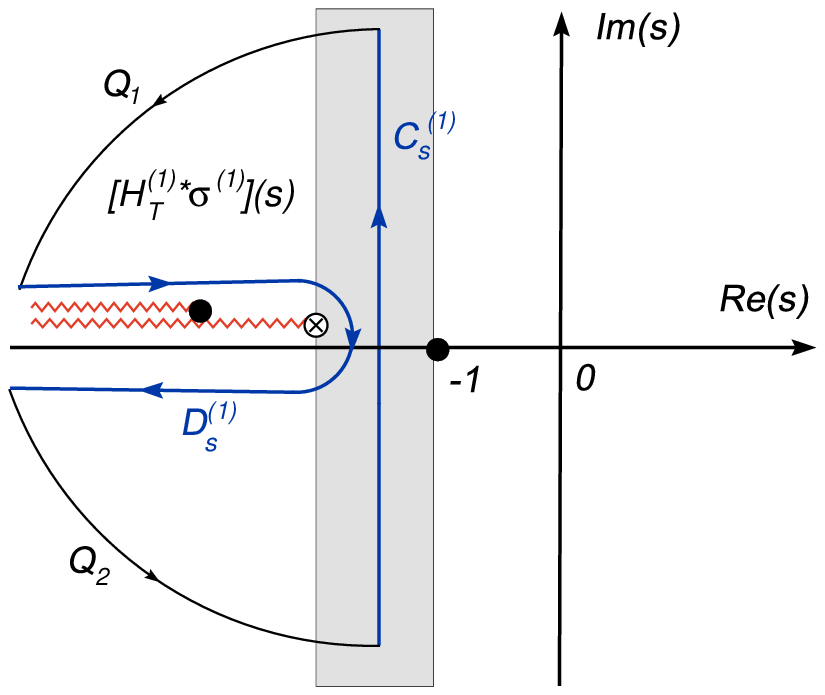}
\end{center}
\caption{ 
The singularity structure of $\sigmatilde^{(0)}(x,s)\tilde H_T(-s,Q^2)$ and
$\sigmatilde^{(1)}(x,s)\tilde H_T(-s,Q^2)$,
and the integration contours in the complex $s$-plane. 
The meaning of all symbols is as in Fig.\ \ref{fig:mel0} --
additionally, pieces of the closed integration contour are shown:
${\cal C}^{(1)} _s,\,{\cal D}^{(0)}_s,\,{\cal D}^{(1)}_s$, etc.
\label{fig:meldec}
}
\end{figure}

The singularities of the Mellin transform 
$\tilde\sigma^{(n)}(x,s)$ emerging at the branch point to the left of 
its fundamental strip are the same as the corresponding singularities of 
the functions: $\tilde\sigma^{(0)}(x,s),\,\tilde\sigma^{(1)}(x,s), \ldots, \tilde\sigma^{(n-1)}(x,s)$.
The functions $H^{(n)}_{T,L}(\rhat)$ are shifted 
with respect to $H_{T,L}(\rhat)$ by a finite power series in $1/\rhat^2$, 
so their Mellin transforms are identical and equal to the Mellin transform
$\Htilde_{T,L}(s,Q^2)$ for all $n$.
The series subtraction results only in the already discussed
shift in the position of the fundamental Mellin strip.
Therefore, the difference 
\beq
\Delta^{(n)}[\sigma H_{T,L}](x,Q^2) \,= \int_{{\cal C}_s ^{(n-1)}} \,
{ds \over 2\pi i} \, \tilde\sigma^{(n-1)} (x,s) \, 
\tilde H^{(n-1)}_{T,L}(-s,Q^2)  \; - \;  
\int_{{\cal C}_s ^{(n)}} \,
{ds \over 2\pi i} \, \tilde\sigma^{(n)} (x,s) \, 
\tilde H^{(n)}_{T,L}(-s,Q^2)
\label{eq:masters}
\eeq
defines the contribution of the $n$-th singularity (i.e.\ the cut 
discontinuity with the branch point at $s=-n$) to the integral in
Eq.\ (\ref{eq:parsival}), see Fig.\ \ref{fig:meldec} for  illustration. 
In our analysis we identify this contribution with the $\tau=2n $ twist component 
of $\sigma^{\gamma^*p}_{T,L}$. 

The Mellin integrals in Eq.\ (\ref{eq:masters}) 
may be expressed in the coordinate space to give a formula
that facilitates a direct determination of all twists in the coordinate
representation: 
\beq
\sigma^{(2n)} _{T,L}(x,Q^2) \; = \;
\int_0 ^{\infty} {dr^2 \over r^2} \left\{
\sigma^{(n-1)}(x,r)\,  H^{(n-1)}_{T,L}(rQ) \, - \,
\sigma^{(n)}(x,r)\,  H^{(n)}_{T,L}(rQ)\right\}.
\label{eq:masterr}
\eeq
Clearly, the twist decomposition would be complete and
\beq
\sum_{n=1} ^{\infty} 
\sigma^{(\tau=2n)} _{T,L}\,=\,
\sigma^{\gamma^*p}_{T,L}
\eeq
provided that the infinite summation of 
$\Delta^{(n)}[\sigma H_{T,L}](x,Q^2)$ is convergent\footnote{In fact the
series in not convergent; the expression was obtained assuming the
validity of the asymptotic $H_{T,L}(rQ)$ expansion for large $rQ$ 
down to $rQ = 0$. Therefore, the obtained series is asymptotic.}.

The prescription given by Eq.\ (\ref{eq:masterr}) may be also applied to
the original GBW dipole cross section which Mellin transform has a series of isolated 
poles at $s=-n$ instead of the series of cut singularities. 
In this case, in Eq.~(\ref{eq:sigsub}) 
a polynomial is subtracted and the Mellin transform of $\sigma^{(n)}(x,r)$ is
identical to $\tilde\sigma(x,s)$ given by Eq.~(\ref{eq:melGBW}). 
Therefore,  formulae 
(\ref{eq:masters}) and (\ref{eq:masterr}) may also  be applied to extract 
the contribution from all singularities of $\tilde\sigma(x,s)  H_{T,L}(-s,Q^2)$
to give the twist expansion in the case of the GBW dipole cross section. 
An explicit numerical check showed that the expansion obtained using 
prescription (\ref{eq:masterr}) agrees with the analytic
results in \cite{Bartels:2000hv}.

\section{Heavy quarks}
\label{sec:8}

So far we studied the massless quark contribution. 
Within the $k_T$-factorization approach it is straightforward
to study also the case with a non-zero quark mass. 
In particular, the Mellin transforms of the photon wave functions 
squared  with  $m_f \ne 0$, which generalize expressions (\ref{eq:a7}) and (\ref{eq:a7l}),  
are known \cite{GolecBiernat:1998js}.
We shall denote them by $\tilde H_{T,L}(s,Q^2,m_f^2)$.
The contribution of a heavy quark to the $\gamma^* p$ 
cross section may be obtained using 
the Parseval formula (\ref{eq:parsival}) in which the replacement 
$\tilde H_{T,L}(-s,Q^2) \,\to\, \tilde H_{T,L}(-s,Q^2,m_f^2)$ 
is made. For $\mathrm{Re}\,s >0$, the functions $\tilde H_{T,L}(s,Q^2,m_f^2)$ 
are regular in $s$. Therefore, the $s$-singularity structure of the
integrand $\tilde\sigma(x,s)\tilde H_{T,L}(-s,Q^2,m_f^2)$
in (\ref{eq:parsival}) is determined by the singularity structure
of $\tilde\sigma(x,s)$. Hence, for heavy quarks the twist-$\tau$ 
component is determined by the $n$-fold scattering component of the
dipole cross-section,
\beq
\sigma^{(\tau=2n)} _{T,L} (x,Q^2,m_f) \;=\; 
\int_0 ^\infty {dr^2 \over r^2}\, H_{T,L}(r^2,Q^2,m_f^2) \, \sigma_n(x,r^2)
\label{eq:theavy}
\eeq
where  $\sigma_n(x,r^2)=\sigma_0 (-1)^{n+1}\, \Omega^n(x,r^2) / n!$.
Note that for heavy quarks with $m_f^2 \gg \mu_0^2$, the integration 
in (\ref{eq:theavy}) does not lead to any infra-red divergences since the 
photon wave function provides an exponential cut-off proportional to $\exp(-r m_f)$ 
for the $r^2$ integration. This was not the case for the light quarks, when
$m_f^2 \ll \mu_0^2$, for which formula (\ref{eq:theavy}) cannot be applied.

\section{Phenomenological results and consequences}
\label{sec:9}

In this section the obtained estimates are presented for the higher twist 
effects in $F_T$, $F_L$ and $F_2$, and also, separately, for the charm quark components of $F_T$, $F_L$ and $F_2$. Additionally, we discuss the phenomenological consequences of our findings for the measurements at the LHC.
We performed an explicit numerical evaluation of higher twist
components of the proton structure functions in the DGLAP improved saturation
model, defined by Eqs.~(\ref{eq:sigmad})--(\ref{eq:sigmad1}) with the evolution scale given 
by Eq.~(\ref{eq:newsatscale}):
\beq\nonumber
\mu^2 \; = \; \left\{   
\begin{array}{ll} 
C/r^2 \;\;\; & \mbox{ for }\;\; r^2 < C/\mu_0^2
\\
\mu_0^2 \;\;\; & \mbox{ for }\;\; r^2 \ge C/\mu_0^2\, .
\end{array} 
\right. 
\eeq
Furthermore, we compared the results to those obtained in \cite{Bartels:2000hv}  
within the GBW model \cite{GolecBiernat:1998js} without the DGLAP evolution. 

The parameters of the DGLAP improved saturation model were fitted to describe all 
HERA data on $F_2$ at $x < 0.01$. In the model we took into account three massless 
quark flavors and the massive charmed quark with $m_c=1.3~{\rm GeV}$. 
The gluon density at the input 
scale $Q_0^2=1~{\rm GeV}^2$  was assumed to take the form 
\be
xg(x,Q^2 _0) = A_g\, x{^{-\lambda}}\,(1-x)^{5.6}\,.
\ee 
The parameters obtained from the best fit with $\chi^2 = 0.94 /{\rm d.o.f}$ are the following: 
\be
C = 0.55\,,~~~~~~~~\mu_0^2=1.62\,,~~~~~~~~
A_g = 1.07\,,~~~~~~~~\lambda=0.14\,,~~~~~~~~\sigma_0 = 22~{\rm mb}\,.
\ee

\subsection{Structure functions}

\begin{figure}
\centerline{\epsfig{file=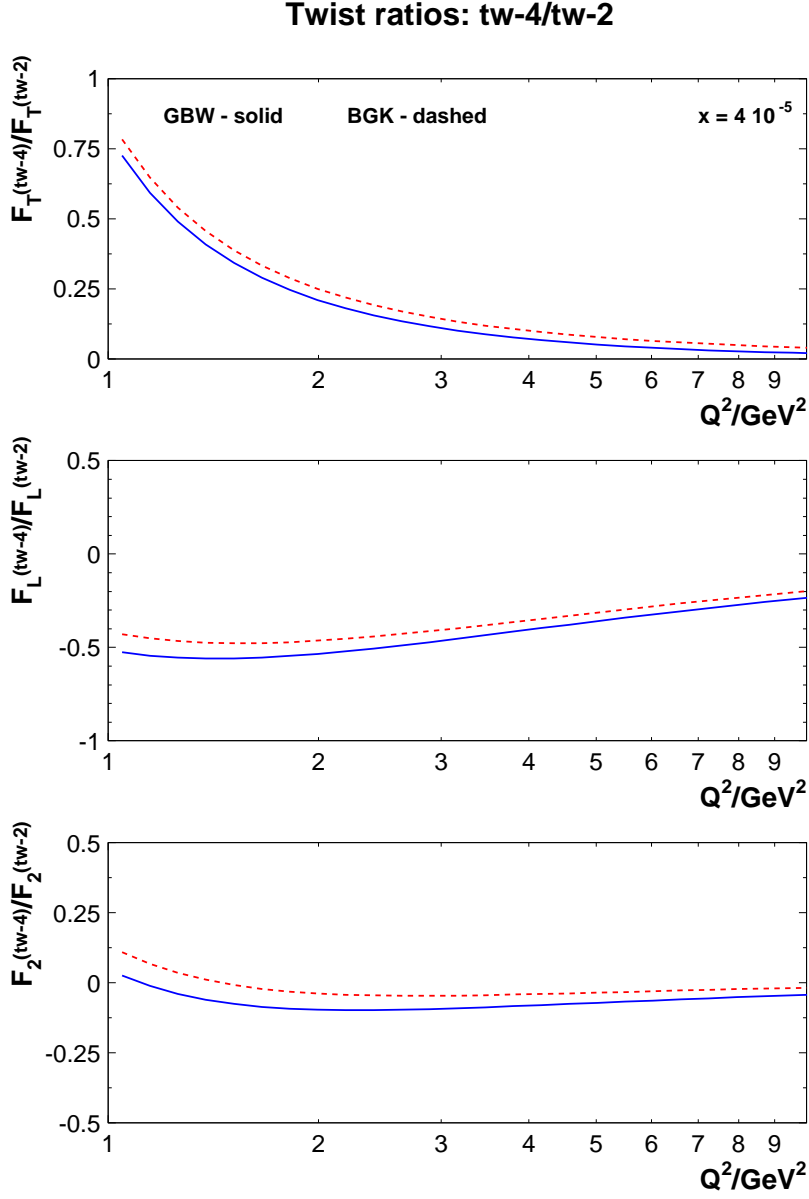,width=0.68\columnwidth}}
\caption{
The ratio of twist-4 to twist-2 components of 
$F_T$, $F_L$ and $F_2$ at $x=4\cdot 10^{-5}$ in the GBW model with charm
(continuous lines) and in the DGLAP improved saturation model (dashed lines).
\label{fig:twist}
}
\end{figure}

The obtained relative twist-4 corrections (with respect to the twist-2 ones) 
to the structure functions $F_T$, $F_L$ and $F_2$ are displayed in Fig.~\ref{fig:twist}, as a function of $Q^2$, for $x=4\cdot 10^{-5}$ (for this value the saturation scale  $Q_s(x)=1~{\rm GeV}$ in the GBW model with charm). The continuous curves obtained in \cite{Bartels:2000hv} correspond to the GBW model with charm quarks~\cite{GolecBiernat:1998js}, and the dashed ones are obtained in the DGLAP improved saturation model (BGK)~\cite{Bartels:2002cj} with the parameters given above.
The differences between the GBW model and the BGK models are visible, but rather small.
The qualitative picture is fully consistent between the models and agrees very well with
the results of the analytic analysis outlined in Sec.~\ref{sec:6}. 
Thus, the higher twist corrections are strongest in $F_L$, and much 
weaker in $F_T$. In $F_2$ there occurs a rather fine cancellation between 
the twist-4 contributions to $F_T$ and $F_L$, at all $Q^2$, 
down to 1~GeV$^2$. Although an effect of this kind was expected, it still
remains somewhat surprising that this cancellation works so well.

\begin{figure}
\centerline{\epsfig{file=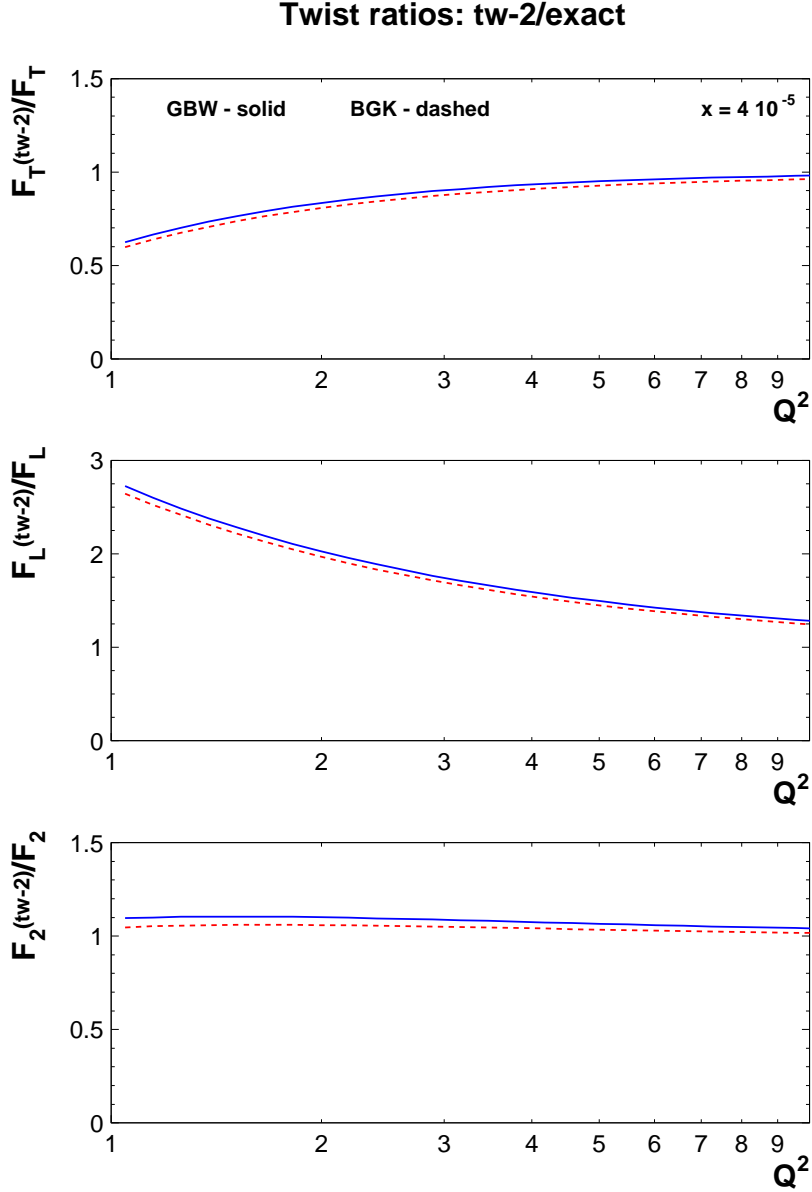,width=0.68\columnwidth}}
\caption{
The ratio of the twist-2 component to the full light quark result of the
model for $F_T$, $F_L$ and $F_2$ at $x=4\cdot 10^{-5}$ in the GBW model with charm
(continuous lines) and in the DGLAP improved saturation model (dashed lines).
\label{fig:t2tototal}}
\end{figure}

We also show in Fig.~\ref{fig:t2tototal} the ratio of the twist-2 component to the full dipole model result for $F_T$, $F_L$ and $F_2$.
The full result incorporates the resummed contributions of all twists. On the qualitative level, the effect of all higher twists shown in Fig.~\ref{fig:t2tototal} are similar to the effect of twist-4 in Fig.~\ref{fig:twist}, indicating that the higher twist corrections are driven by the twist-4 contribution down to $Q^2=1$~GeV$^2$. It is remarkable, that the cancellation of the higher twist effects in $F_2$ occurs also in the all-twist result. Clearly, the results shown in Fig.~\ref{fig:t2tototal} indicate that at lower $Q^2$, the conventional twist-2 calculations underestimate the value of $F_T$, significantly overestimate the value of $F_L$ and slightly overestimate the value of $F_2$. 

One should stress, that the theoretical conclusions about the strength of the higher twist corrections, related to the powers of $\alpha_s$, are only valid in the perturbative range, where $\alpha_s$ is small. They are therefore, well justified for $Q^2$ above, say, 5~GeV$^2$. In that region, indeed, the difference between higher twists in $F_L$ and $F_T$ is quite pronounced. At lower $Q^2$, where $\alpha_s$ is not small, the differences in powers of $\alpha_s$ should not lead to quantitatively distinct results in $F_T$ and $F_L$. Indeed, at $Q^2=1$~GeV$^2$, the relative twist-4 corrections to $F_T$ and $F_L$ are 30\% and 50\% correspondingly, that is they do not differ very much.

\subsection{Charm contribution}

\begin{figure}
\centerline{\epsfig{file=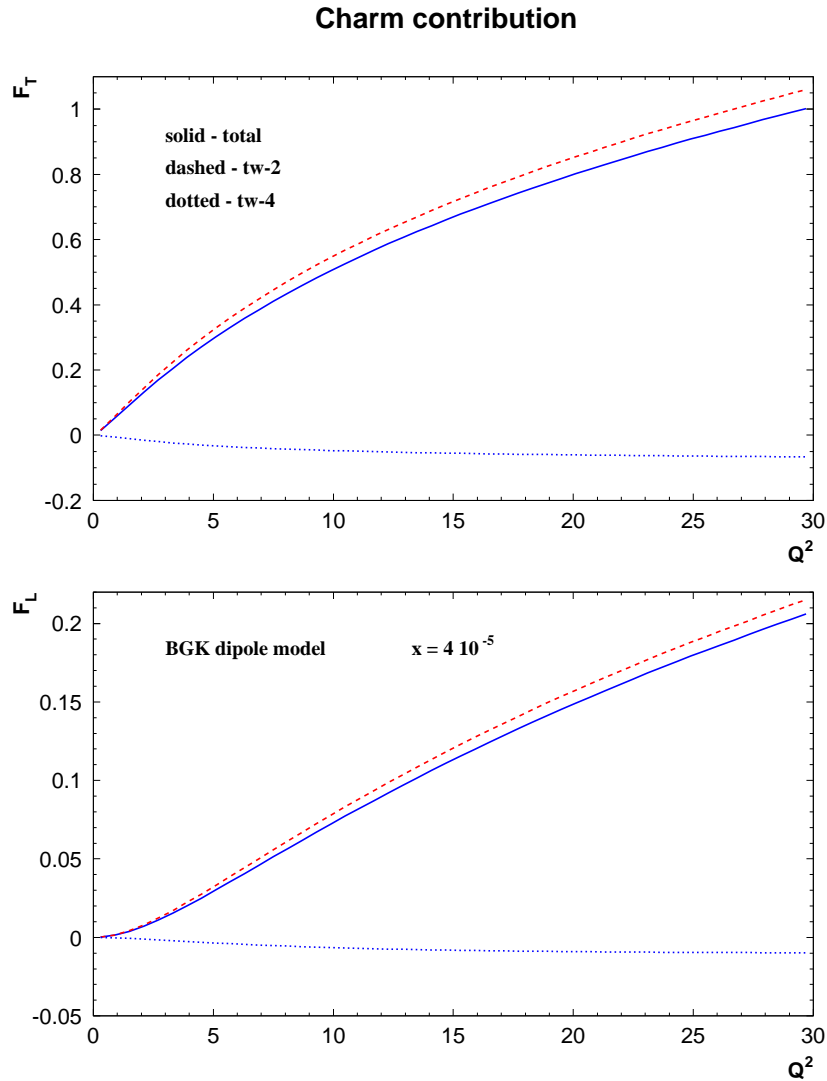,width=0.68\columnwidth}}
\caption{
Charm contribution to $F_T$ and  $F_L$ at $x=4\cdot 10^{-5}$ in the DGLAP improved saturation model. Shown are the contributions of the twist-2 component (dashed lines), the twist-4 component (dotted lines) and the full result (continuous lines).
\label{fig:charm}}
\end{figure}

The DIS cross-section at small-$x$ and a moderate $Q^2$ receives significant 
contribution from the charmed quark. The contribution of the bottom quark may be safely neglected due to its small charge e$_b = 1/3$ and its large mass. 
For the region of interest, $Q^2 \sim 10$~GeV$^2$, the mass, $m_c$, of the charmed quark cannot be neglected, as $Q^2 \sim 4m_c^2$. Therefore, our extraction of the higher twist effects in charm structure functions $F_T ^c$ and $F_L ^c$ is based on the results of Sec.~\ref{sec:8}. The results for higher twist effects in the charm structure functions $F^c _T$ and $F^c _L$ are shown in Fig.~\ref{fig:charm}. Displayed are the twist-2 and twist-4 components and the all-twist result. In contrast with the case of the light quarks, the higher twist effects introduce negative corrections both in $F^c _T$ and in $F^c _L$, and the magnitude of the ratio of twist-4 to twist-2 contributions is similar in both cases and reaches a few percent. Consequently, the effect of higher twists in $F^c _2$ is similar.

\subsection{Comparison with $F_L$ data}

\begin{figure}
\centerline{\epsfig{file=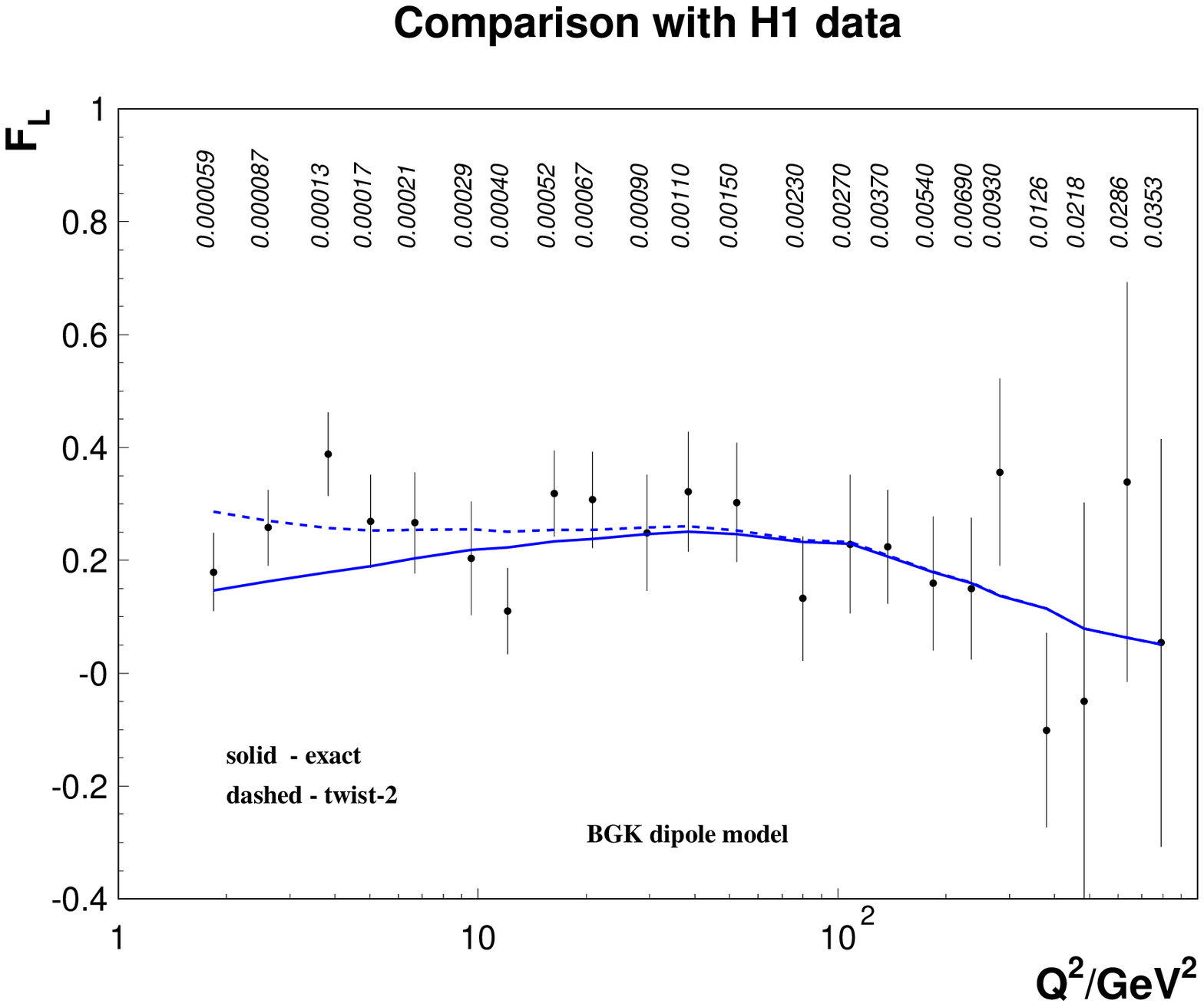,width=0.69\columnwidth}}
\caption{
Comparison of the DGLAP improved saturation model with the preliminary H1 on $F_L$ \cite{FLH1new}.
The solid line is the all-twist result while the dashed line shows the twist-2 contribution. The value of~$x$ is indicated for each data point.
\label{fig:H1data}}
\end{figure}

Recently, new measurements were performed of the proton $F_L$ structure function in a wide kinematic range~\cite{FLH1new}. The measurements probe $F_L$ for correlated $(x,Q^2)$ pairs down to $x=5.9 \cdot 10^{-5}$ and $Q^2=1.8$~GeV$^{2}$, see Fig.~\ref{fig:H1data}. The data in the lowest range of $x$ are particularly interesting, as in this region, the leading twist, fixed order DGLAP calculations face intrinsic problems~\cite{ThoW,MSTW}. Specifically, in that region, the convergence of the subsequent $F_L$ approximations within perturbative expansions is rather poor, up to the next-to-next-leading order (NNLO) approximation~\cite{MSTW}. In addition, at very small~$x$ and low~$Q^2$, the estimated $F_L$ becomes negative, violating the fundamental condition of positivity~\cite{MSTW}. This indicates that the DGLAP treatment in this region has to be improved. In  what follows, we shall present the comparison of the new $F_L$ data with the dipole model results, and we shall shortly compare our approach with another successful approach to $F_2$ and $F_L$, based on the leading twist DGLAP scheme, improved by a small~$x$ resummation~\cite{ThoW}.

In Fig.~\ref{fig:H1data} we show the comparison of our results with the preliminary data on $F_L$ from the H1 collaboration. In the top of the plot, the values of $x$ are indicated for each data point. Note that the experimental data points show strong correlation between the values of $Q^2$ and $x$. Thus, small $Q^2$ values are measured for smaller values of $x$. The solid curve represents the all twist result from the DGLAP improved saturation model applied in this paper,  while the dashed line shows the twist-2 contribution within this model. The difference between the two curves comes from the negative higher twist terms, with a dominant contribution of twist-4. The description of the data provided by the model is good, both for the twist-2 approximation and the all-twist result. We stress, that all the model parameters are fixed by the fit to $F_2$ data and no new parameters are introduced in the description of $F_L$.

Clearly, the low $Q^2$ region of the plot, where~$x$ is also small, is highly sensitive to higher twist effects. In particular, for the lowest measured values of $(x,Q^2)$, the twist-2 contribution is already about two times larger then the exact result. Unfortunately, the current experimental errors are sizable and no evidence for higher twist effects can be drawn from the measurements, yet. We stress, however, that $F_L$ at small~$x$ and $Q^2$ should be an excellent observable to find such effects, provided that the experimental errors may be further reduced.

The defects of the fixed order DGLAP description of $F_L$ at small~$x$ and~$Q^2$ were shown to be partially cured by including into the DGLAP framework a resummation of small~$x$ corrections, enhanced by powers of $\log x$. The resummation, proposed by Thorne and White (TW)~\cite{ThoW}, absorbs the NLL~BFKL effects at the leading twist into the NLO DGLAP evolution. In the currently relevant kinematic range, the description of $F_L$ based on the TW approach is remarkably similar~\cite{MSTW} to the one obtained within a saturation model with the impact parameter dependence (the, so called, b-Sat model)~\cite{bSat}. In addition, the b-Sat model results for $F_L$ agree well with the results of this paper. 
The TW scheme provides a good description of the existing $F_L$ data. The $F_L$ at small $x$ and $Q^2$ following from the saturation models is significantly lower than the corresponding TW result, but the differences are not pronounced enough to permit a discrimination between the approaches with the present data. Let us, however, stress, that the asymptotic $x\to 0$ (or $Q^2 \to 0$) behaviour of the structure functions should be different in approaches consistent with unitarity constraints, (as e.g.\ the saturation models), and the leading twist approach. In the former case, $F_L$ should vanish in the limit, while in the latter case it should remain non-zero. Thus, one expects, that the leading twist approach should be insufficient at a very low~$x$ and fixed $Q^2$, and that the inclusion of higher twist effects should be necessary in that limit.

\subsection{Discussion and implications for the LHC}

The analysis performed in this paper shows that the importance of the higher twist corrections may essentially depend on the process and the probe. In particular, higher twist effect in the structure function $F_2$ are strongly suppressed due to rather fine cancellations between $F_T$ and $F_L$. Such cancellations are not expected to occur in a generic case. For instance, the higher twist effect in $F_L$ are enhanced. Thus far, parton density functions (pdfs) in DIS were fitted mostly to the $F_2$ data. Due to small higher twist effects in $F_2$, one expects that the suppressed higher twist contributions should not affect the quality of the determination of pdfs. This is, certainly, a good news. The estimated correction due to higher twist effects in $F_L$ at small~$x$ and a moderate $Q^2$ is, however, much larger, and this correction should be taken into account when including the $F_L$ data into fits of pdfs.

The example of $F_2$ and $F_L$ in the DIS shows that the multiple scattering (higher twist) effects are probed in various ways, depending on the observable. Similar differences in the magnitude of higher twist effects in various observables may occur in the hadronic collisions, e.g.\ in $pp$ collisions at the LHC. In particular, cancellations present in $F_2$ is not expected for the bulk of LHC observables probing the gluon distributions at small-$x$. Thus, in general, the relative effects of higher twists at the LHC should be larger than they are in $F_2$.  As an example, let us give the case of the forward Drell-Yan process, that can be effectively described using the dipole formulation~\cite{FDY}. At LHC-b, the Drell-Yan process may be probed at moderate $Q^2 \sim 10$~GeV$^2$ and $x \sim 10^{-6}$, what should provide a gold-plated probe of the gluon density at small~$x$. However, the higher twists effects may be quite strong there. In particular, let us invoke an example of the Lam--Tung relation~\cite{LT} that holds for angular distribution of Drell-Yan lepton pairs. According to this relation, the twist-2 contribution to one of the angular components of the dilepton distribution vanishes in the leading logarithmic approximation. Therefore, higher twist effects in this component should be enhanced, in analogy to the case $F_L$~\cite{LT,HTLT}. Thus, given the low values of $x$ and $Q^2$, that can be reached in the measurements in LHC-b, the violations of Lam-Tung relation should provide a sensitive probe of the higher twist gluonic operators at small~$x$. On the other hand, the higher twist effect may be also large in the total cross-section of the forward Drell-Yan process. In that case, a determination of gluon density at small~$x$, based on the leading twist contribution alone would be inaccurate,  and the higher twist contributions should be taken into account.
Besides that, the higher twist effects may be larger in processes with gluons, like e.g.\ the forward gluonic jet productions, where the multiple scattering of the gluon is enhanced by its color charge, as compared to the quark rescattering in the DIS case. In such processes, we do not expect that any cancellations of rescattering effects should occur, of the type found in $F_2$.

\section{Conclusions}\label{sec:10}

In this paper the leading higher twist contributions to proton structure functions, $F_2$, $F_T$ and $F_L$, at small Bjorken~$x$ and moderate $Q^2$ were analyzed. The problem was analyzed theoretically confronting two different approaches.
In the fist approach, we focused on a subset of QCD diagrams describing contributions of quasipartonic gluon operators, that should dominate the higher twist effects in the deeply inelastic scattering at small~$x$.  We demonstrated, that this subclass of the diagrams, at the leading logarithmic approximations and in the large $N_c$ leads to a picture consistent with the DGLAP improved saturation model. In contrast, we considered also the problem of higher twists in the Balitsky-Kovchegov framework, in which, the BFKL Pomeron fan diagram are resummed. In this approach, higher twist contributions coming from the fan diagrams vanish in the leading $\log Q^2$ approximation.

The pattern of the most important twist-2 and twist-4 contributions to  $F_2$, $F_T$ and $F_L$ is determined by the properties of the quark loop through which the virtual photon interacts with the gluonic field of the proton. Therefore, it is universal and its key features should not depend on the model details. Those features are: (i) the twist-4 correction to $F_T$ enters only at the NLO, and so, the twist-4 correction to $F_T$ is suppressed; (ii) the twist-2 contribution to $F_L$ enters at NLO, and the LO twist-4 term in $F_L$ is relatively enhanced and more important; (iii) the relative sign of twist-4 corrections to $F_T$ and $F_L$ is opposite, and the higher twist effects partially cancel in $F_2 = F_T + F_L$. These general conclusions were then confirmed by a quantitative phenomenological analysis.

We performed a numerical twist analysis of the DIS cross-sections at small-$x$ within the DGLAP improved saturation model. In order to carry out a quantitative estimate of the higher twist effects in the structure functions, we proposed a method allowing for a direct, numerical twist decomposition of the saturation model cross sections. The method was then applied to the DGLAP improved saturation model, fitted to the HERA $F_2$ data. Contributions of twist-2, twist-4 and all twists to $F_2$, $F_T$ and $F_L$ were then extracted. We found a strikingly good cancellation of the higher twist effects in $F_2$, for which, at $x=3\cdot 10^{-4}$, the relative correction from higher twists is found to be at a few percent level down to $Q^2=1$~GeV$^2$. The higher twist corrections to $F_T$ were found to be moderate, below 10\% for $Q^2>3$~GeV$^2$ at $x=3\cdot 10^{-4}$. On the other hand, the twist-4 correction in $F_L$ was found to be large, about 50\%, at $Q^2=1$~GeV$^2$, and still sizeable, about 20\%, at $Q^2=10$~GeV$^2$. Therefore, whereas the leading twist analyses of $F_2$ are fully justified, one should include the higher twist effects in analyses of the $F_L$ data at small~$x$ and moderate $Q^2$. We also found that the saturation model description of the recent $F_L$ measurements at small~$x$ and low~$Q^2$ is good.
Unfortunately, the data are not precise enough to prove that the inclusion of higher twist corrections improves the description of the data.

Finally, some implications were discussed of the results for analyzes of the LHC data.
In particular, we stressed a strong process-dependence of the higher twist contributions, exemplified before by the striking differences between $F_2$ and $F_L$. It follows from our analysis, that $F_2$ is protected by cancellations from the higher twist effects, and such cancellations are not expected to be generic. Therefore, the higher twist effects in some LHC observables may be much stronger that they are in $F_2$. Thus, it is crucial to estimate higher twist effects when attempting a precise determination of parton densities in LHC measurements at small~$x$ and moderate $Q^2$, like e.g.\ in the forward Drell-Yan process at low $Q^2$, or in the forward jet production.


\section*{Acknowledgements}
LM acknowledges the support of the DFG grant SFB~676.
This work is partially supported by the grant MNiSW no. N202 249235.


\appendix

\section{Mellin transforms of $H_{T,L}$}
\label{app:a}

Let us compute the Mellin transform of 
$H_T(rQ)$  given by Eq.~(\ref{eq:3a})
\be\label{eq:a1}
\Htilde_{T}(s,Q^2)\,= \int_0^\infty dr^2\,(r^2)^{s-1}\,H_{T}(rQ)\,.
\ee
Substituting Eq.~(\ref{eq:3a}) we obtain
\be
\Htilde_{T}(s,Q^2)\,= \left(\frac{Q^2}{4}\right)^{-s} \Htilde_{T}(s)
\ee
where
\be\label{eq:a2}
\Htilde_{T}(s)=4^{-s}A_0
\intlim_0^1 dz\,[z^2+(1-z)^2]\,z(1-z)
\intlim_0^\infty d\rhat^2\,(\rhat^2)^{s}\,K_1^2(\sqrt{z(1-z)}\, \rhat)\,.
\ee
with $\rhat=rQ$.
Changing the variable, $y^2=z(1-z)\rhat^2$, we find
\be\label{eq:a3}
\Htilde_{T}(s)=4^{-s}A_0
\intlim_0^1 dz\,\frac{z^2+(1-z)^2}{z^s(1-z)^s}
\intlim_0^\infty dy^2\,(y^2)^s\,K_1^2(y)\,.
\ee
The integral over $z$ equals
\be
I_z\equiv\intlim_0^1 dz\,\frac{z^2+(1-z)^2}{z^s(1-z)^s}=
2\intlim_0^1 dz\,z^{2-s}(1-z)^{-s}=2\,\frac{\Gamma(3-s)\Gamma(1-s)}{\Gamma(4-2s)}
\ee 
where we used the definition of the Euler beta function. The gamma function in the denominator
can be written as
\be\label{eq:a4}
\Gamma(2(2-s))=\frac{2^{2(2-s)-1}}{\sqrt{\pi}}\,\Gamma(2-s)\Gamma(2-s+1/2)
\ee
and from this we have
\be\label{eq:a5}
I_z=\frac{\sqrt{\pi}}{4^{1-s}}\,
\frac{\Gamma(1-s)\Gamma(3-s)}{\Gamma(2-s)\Gamma(5/2-s)}\,.
\ee
The integral over $y^2$ in Eq.~(\ref{eq:a3}) equals
\be\label{eq:a6}
\intlim_0^\infty dy^2\,(y^2)^s\,K_1^2(y)=
2\intlim_0^\infty dy\,y^{2s+1}\,K_1^2(y)=2\,\frac{\sqrt{\pi}}{4}\,
\frac{\Gamma(2+s)\Gamma(1+s)\Gamma(s)}{\Gamma(3/2+s)}\,.
\ee
Thus, we finally  find 
\be\label{eq:a7}
\Htilde_{T}(s)\,=\,\frac{A_0\pi}{8}\,
\frac{\Gamma(2+s)\,\Gamma(1+s)\,\Gamma(s)\,\Gamma(1-s)\,\Gamma(3-s)}
{\Gamma(3/2+s)\,\Gamma(2-s)\,\Gamma(5/2-s)}\,.
\ee
A similar calculation allows to compute the Mellin transform 
\be
\label{eq:a7l}
\Htilde_{L}(s)\,=\,\frac{A_0\pi}{4}\,
\frac{(\Gamma(1+s))^3\,\Gamma(2-s)}{\Gamma(3/2+s)\,\Gamma(5/2-s)}
\ee

\section{Evolution of  the gluon density}
\label{app:b}

The gluon density used in our analysis,   $g(x,\mu^2)\equiv xG(x,\mu^2)$,
obeys the following  leading logarithmic DGLAP evolution equation:
\beq\label{eq:b1}
\mu^2 {\partial g(x,\mu^2) \over \partial \mu^2} \; = \;
{\alpha_s (\mu^2)\over 2\pi} \int_x ^1 dz \, P_{gg}(z) \,
g\left({x\over z},\mu^2\right)
\eeq
where the contribution from quarks is neglected and 
the gluon splitting function $P_{gg}$ takes the form,
\beq\label{eq:b2}
P_{gg}(z) = 6\left[ {1-z\over z} + z(1-z) + {z\over (1-z)_+} +
{11 \over 12}\delta(1-z)\right] -{n_f \over 3}\delta(1-z)
\eeq
and the one loop strong coupling is given by
\beq\label{eq:b3}
\alpha_s (\mu^2) = {b_0 \over \log(\mu^2 / \Lambda^2)}
\eeq
with $b_0 = 12\pi / (33 - 2n_f)$.
This equation may be solved in the double  Mellin moment space,
\beq\label{eq:b4}
g(x,\mu^2) = \int {d\omega \over 2\pi i}\;x^{-\omega}\! \int {d\gamma \over 2\pi i} \;
{\tilde g}(\omega,\gamma) \left({\mu^2 \over \Lambda ^2}\right)^\gamma 
\eeq
where ${\tilde g}(\omega,\gamma)$ obeys the following equation
\be\label{eq:b5}
-{\partial \over \partial \gamma}\left\{
\gamma \tilde g(\omega,\gamma)\right\} =
{b_0 \over 2\pi} \, \tilde P_{gg} (\omega) \,  \tilde g(\omega,\gamma)
\ee
and the splitting kernel in the Mellin representation, $\tilde P_{gg} (\omega)$, is given by
\beq\label{eq:b6}
\tilde P_{gg} (\omega) = \int_0^1 dz\, z^{\omega} P_{gg}(z) =
6 \left[
{1\over \omega} - {1 \over \omega+1} +   
{1 \over \omega+2} - {1 \over \omega+3} - \gamma_E - \psi(\omega+2)
\right] + {33 - 2n_f \over 6}\,. 
\eeq
The general solution to  Eq.~(\ref{eq:b5}) reads
\beq
\tilde g(\omega,\gamma)\, =\, \tilde g_0(\omega) \, 
\gamma^{-1-{b_0\over {2\pi}} \tilde P_{gg}(\omega)}
\label{eq:gluon}
\eeq
where  $\tilde g_0(\omega)$ is an arbitrary function which may be fixed
using an initial condition. Thus, the solution expressed in terms of the 
original  variables $(x,\mu^2)$ is given by
\beq\label{eq:b8}
g(x,\mu^2)\, = 
\int {d\omega \over 2\pi i}\;x^{-\omega}\,\tilde  g_0(\omega)\!  
\int {d\gamma \over 2\pi i}\, 
\gamma^{-1-{b_0 \over 2\pi}  \tilde P_{gg} (\omega) }
\left({\mu^2 \over \Lambda^2}\right)^\gamma. 
\eeq
The contour integral over $\gamma$ may be performed for all 
$\mu^2 > \Lambda^2$ after the integration contour is deformed to envelope the cut 
along the negative real half-axis. We find 
\be\label{eq:basic}
g(x,\mu^2)\, = \int {d\omega \over 2\pi i}\;x^{-\omega}\,\tilde g(\omega,\mu^2) 
\ee
where
\beq
\tilde g(\omega,\mu^2)\, =\, 
{\tilde g_0(\omega) \over  
\Gamma \left(1+{b_0 \over 2\pi} \tilde P_{gg}(\omega) \right)} \
\; 
\left[\log\left({\mu^2 \over \Lambda^2}\right)\right]
^{{b_0 \over 2\pi} \tilde P_{gg} (\omega)}.
\label{eq:ghat}
\eeq
The initial condition for the DGLAP equation at some scale 
$\mu_0^2\gg \Lambda^2$ is given  by its Mellin transform $\tilde g(\omega,\mu_0^2)$. 
Thus, writing (\ref{eq:ghat}) for $\mu^2 = \mu_0^2$, we obtain
\beq
\tilde g_0(\omega) \; = \; \tilde g(\omega,\mu_0^2)  \;
\Gamma \left(1+{b_0 \over 2\pi} \tilde P_{gg}(\omega) \right)\;
\left[\log\left({\mu_0^2 \over \Lambda^2}\right)\right]
^{-{b_0 \over 2\pi} \tilde P_{gg} (\omega)}\; ,
\label{eq:g0}
\eeq
which in turn, after the substitution to (\ref{eq:basic}),  leads to the well known form 
\beq\label{eq:b11}
g(x,\mu^2) = \int {d\omega \over 2\pi i}\; 
x^{-\omega} \;\tilde g(\omega,\mu_0^2) 
\left[ {
\log(\mu^2 / \Lambda^2) \over \log(\mu_0^2 / \Lambda^2) }
\right]^{{b_0 \over 2\pi}\tilde P_{gg}(\omega)}.
\eeq

We also need the double Mellin representation
of the product $\alpha_s(\mu^2)g(x,\mu^2)$. 
In the mixed $(\omega,\mu^2)$ representation the DGLAP equation (\ref{eq:b1}) reads
\beq\label{eq:b12}
\mu^2 {\partial  \tilde  g(\omega,\mu^2)\over \partial \mu^2}\; = \;
{\tilde P_{gg}(\omega)\over 2\pi} \,   \alpha_s(\mu^2)\, \tilde g(\omega,\mu^2)\,. 
\eeq
Taking the Mellin moment (\ref{eq:mellinmu}) of both sides (with $s=\gamma$), we obtain
\be
\gamma\,\tilde g(\omega,\gamma)\,=\,
{\tilde P_{gg}(\omega)\over 2\pi} \,\widetilde{\alpha_s g}(\omega,\gamma)\,.
\ee
Thus,   after inserting relation (\ref{eq:gluon}) we find
\be\label{eq:asgluon}
\widetilde{\alpha_s g}(\omega,\gamma)
\,=\,\frac{2\pi\tilde g_0(\omega)}{\tilde P_{gg}(\omega)}\,\,
\gamma^{-{b_0\over {2\pi}} \tilde P_{gg}(\omega)}
\ee
which after coming back to the $x$ variable reads 
\beq
\label{eq:asgluon1}
\widetilde{\alpha_s g}(x,\gamma) \; = \;
\int {d\omega \over 2\pi i}\, x^{-\omega}\;
\frac{2\pi\tilde g_0(\omega)}{\tilde P_{gg}(\omega)}\,\,
\gamma^{-{b_0\over {2\pi}} \tilde P_{gg}(\omega)}\,.
\eeq
Both functions  have logarithmic cut singularity
in the complex $\gamma$-plane with the branch point at $\gamma=0$.


\end{document}